\documentclass[manuscript]{acmart} %,review,anonymous

\usepackage{tcolorbox}
\usepackage{listings}
\usepackage{xcolor}

\AtBeginDocument{%
  }

\setcopyright{acmlicensed}
\begin{document}

\newcommand{\myListItem}[1]{\textbullet~{\textbf{#1}}}
\newcommand{\rewriteLater}[1]{{#1}}

%%
%% The "title" command has an optional parameter,
%% allowing the author to define a "short title" to be used in page headers.
\title{Can You Move These Over There? An LLM-based VR Mover for Supporting Object Manipulation}

%%
%% The "author" command and its associated commands are used to define
%% the authors and their affiliations.
%% Of note is the shared affiliation of the first two authors, and the
%% "authornote" and "authornotemark" commands
%% used to denote shared contribution to the research.

\author{Xiangzhi Eric Wang}
\affiliation{%
  \institution{The Hong Kong Polytechnic University}
  \country{Hong Kong SAR}
}

\author{Zackary P. T. Sin}
\affiliation{%
  \institution{The Hong Kong Polytechnic University}
  \country{Hong Kong SAR}
}

\author{Ye Jia}
\affiliation{%
  \institution{The Hong Kong Polytechnic University}
  \country{Hong Kong SAR}
}

\author{Daniel Archer}
\affiliation{%
  \institution{University College London}
  \country{United Kingdom}
}

\author{Wynonna H. Y. Fong}
\affiliation{%
  \institution{Heep Yunn School}
  \country{Hong Kong SAR}
}

\author{Qing Li}
\affiliation{%
  \institution{The Hong Kong Polytechnic University}
  \country{Hong Kong SAR}
}

\author{Chen Li}
\affiliation{%
  \institution{The Hong Kong Polytechnic University}
  \country{Hong Kong SAR}
}

%%
%% By default, the full list of authors will be used in the page
%% headers. Often, this list is too long, and will overlap
%% other information printed in the page headers. This command allows
%% the author to define a more concise list
%% of authors' names for this purpose.
% \renewcommand{\shortauthors}{Trovato et al.}

%%
%% The abstract is a short summary of the work to be presented in the
%% article.
\begin{abstract}
In our daily lives, we can naturally convey instructions for the spatial manipulation of objects using words and gestures. Transposing this form of interaction into virtual reality (VR) object manipulation can be beneficial. We propose \textit{VR Mover}, an LLM-empowered solution that can understand and interpret the user’s vocal instruction to support object manipulation. By simply pointing and speaking, the LLM can manipulate objects without structured input. Our user study demonstrates that \textit{VR Mover} enhances user usability, overall experience and performance on multi-object manipulation, while also reducing workload and arm fatigue. Users prefer the proposed natural interface for broad movements and may complementarily switch to gizmos or virtual hands for finer adjustments. These findings are believed to contribute to design implications for future LLM-based object manipulation interfaces, highlighting the potential for more intuitive and efficient user interactions in VR environments.

\end{abstract}

%%
%% The code below is generated by the tool at http://dl.acm.org/ccs.cfm.
%% Please copy and paste the code instead of the example below.
%%


%%
%% Keywords. The author(s) should pick words that accurately describe
%% the work being presented. Separate the keywords with commas.
%% A "teaser" image appears between the author and affiliation
%% information and the body of the document, and typically spans the
%% page.
\begin{teaserfigure}
  \centering
  \includegraphics[width=\textwidth]{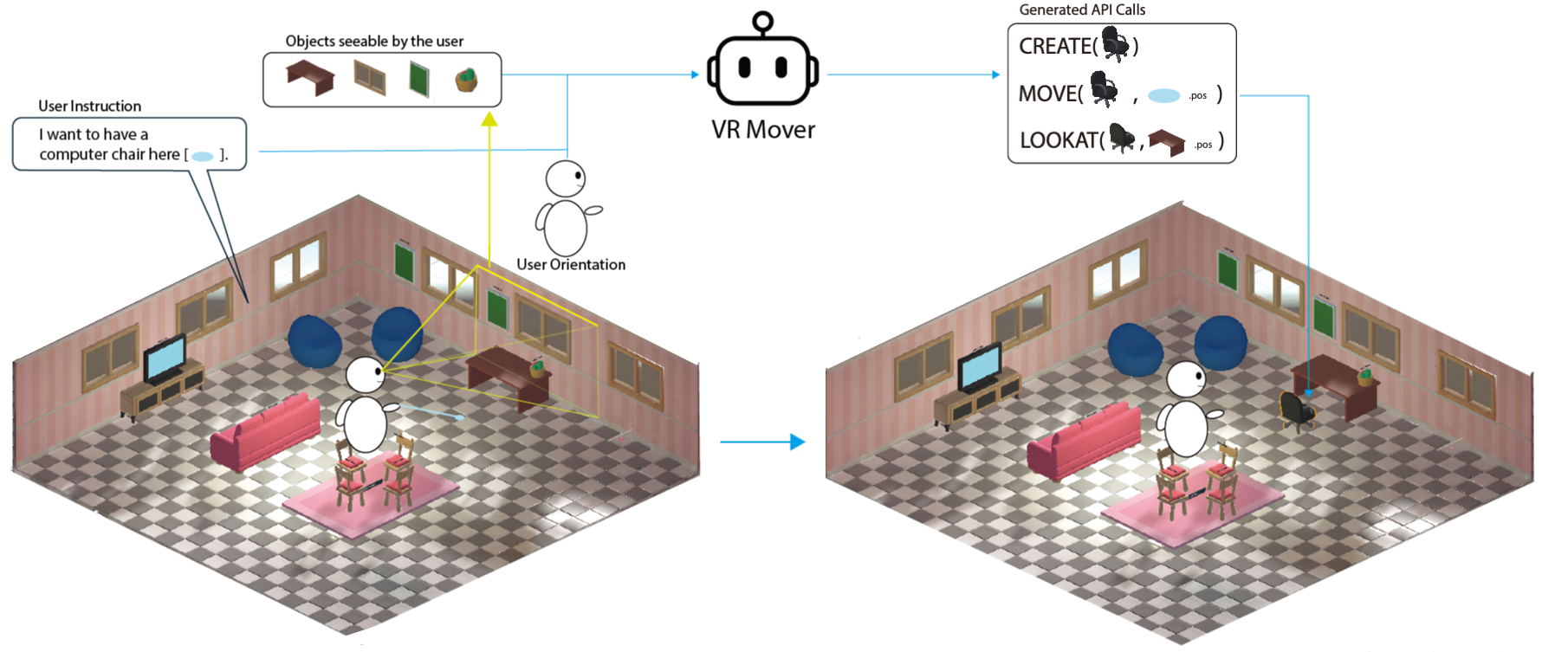}
  \caption{We propose \textit{VR Mover}, an LLM-based interface for supporting object manipulation. It aggregates user-centric information such as what the user is saying, seeing and pointing at to decide how to assist them in the placement of objects.}
  \Description{This figure is showing how the core idea of the proposed VR Mover. There is a before and after room. In which the user's instruction "I want to have a computer chair here" what the user is seeing and the orientation of the user are fed to the VR Mover. Then the VR Mover will generate a set of API calls, which in turns lead to the chair being generated into the after room.}
  \label{fig:teaser}
\end{teaserfigure}

% \received{20 February 2007}
% \received[revised]{12 March 2009}
% \received[accepted]{5 June 2009}

%%
%% This command processes the author and affiliation and title
%% information and builds the first part of the formatted document.
\maketitle

\section{Introduction}
In virtual reality (VR), object manipulation refers to the task of moving and manipulating 3D objects. It is a commonly used task in many applications and is generally associated with scene editing, for example, level design in VR game development or customizing a personal space in the metaverse.
The most common form of object manipulation requires the user to select an object, specify the transform (position, rotation, or scale), and then confirm the maneuver \cite{yu2021gaze}. Despite its importance in VR, this task has been known to induce arm fatigue. This is caused by the so-called "gorilla-arm effect" \cite{hansberger2017dispelling}, which will manifest when the user needs to perform mid-air gestures for an extended period. Another issue is the learning curve. Typically, any interface will involve a learning curve for the user, but for object manipulation, the aforementioned three-step procedure, along with modification of an object's position, rotation, and scale, means most hand-based gestural interfaces have a high barrier to entry. These issues are a concern, as arm fatigue and a rigid interface can affect user experience, and can be a challenge for object-manipulation interface research \cite{yu2021gaze}.

Another issue affecting object manipulation tasks is multi-object manipulation. There are many situations in which the user needs to move multiple objects together. It seems that there is a lack of discussion on how to effectively handle multi-object manipulation, as most research focuses on improving object manipulation in isolation - one object at a time \cite{yu2021gaze}. For handling multiple objects, the assumption seems to be that the user should first select several objects in tandem and then manipulate them as a group. Although this is a valid strategy, this assumes that all of the grouped objects require exactly the same manipulation, which may often not be the case. 

To address the issues raised earlier, we take note of how we perform object manipulation tasks in reality. Specifically, we also perform "scene editing" in the real world, such as when we move residences or decorate rooms. During these processes, it is possible to instruct others to help move objects. In particular, humans tend to combine speech and gestures in the completion of spatial communication tasks \cite{austin2014presentation}. Thus, we combine the use of commanding statements and gestural pointing to provide others with feedback \cite{haviland2000pointing}, and we do so naturally, generally without noticeable arm fatigue or difficulty. We can also give unstructured and context-driven instructions that may also involve multiple objects at once. Since this innate spatial communication is effective and intuitive, our goal is to translate it into VR to create the effect of having a virtual mover who can assist with object manipulation tasks. 

Inspired by how people communicate spatial manipulation instructions in the real world, we propose utilizing a large language model (LLM) to realize a \textit{VR Mover} to support the user's object manipulation. Specifically, the LLM is given several APIs (to move, rotate, scale, create, and remove objects), and based on where the user is pointing and their concomitant instructions, the LLM will decide how to perform the object manipulation task. Because the LLM can deduce context and understand the relationship between current and previous instructions, the user can give more natural instructions. At the same time, the LLM's system prompt is designed in such a way that it understands the spatial arrangement and possible manipulation of objects from the user's perspective. We will later show that the interaction with our \textit{VR Mover} interface is aligned with how human visual working memory operates and can thus, theoretically, be more natural and intuitive to use.

To evaluate our interface, a user study has been prepared with two experimental settings. The first experiment measures how effective an interface is for single mid-air object manipulation and multi-object manipulation. Users will be given several target manipulations to complete. The second experiment provides a free-to-create environment for the users to freely furnish a room. The purpose of this latter part of the study is to see whether the interface is suited to a more practical and creative use case. We compared our LLM-based interface with two other object manipulation interfaces. Primarily, we compared it with a commonly used object manipulation interface where the user can manipulate an object via gizmos \cite{drey2023investigating} and virtual hands \cite{yu2021gaze}, which are prevalent in popular software. Secondly, we compared the LLM-based interface with a voice command variant to highlight the necessity of the LLM. Briefly, the contributions of this paper are as follows:
\begin{itemize}
  \item \textit{VR Mover}, a novel LLM-based object manipulation interface that can handle unstructured, incomplete, and contextualized instructions, and manipulate multiple objects from a user's perspective in real-time. 
  \item The technical implementation of the proposed LLM-based object manipulation interface, and showcasing different ways to use it for object manipulation.
  \item A user study that shows an LLM-based interface like \textit{VR Mover} can improve usability and user experience, as well as reduce arm fatigue and workload. 
\end{itemize}

% ACM's consolidated article template, introduced in 2017, provides a
% consistent \LaTeX\ style for use across ACM publications, and
% incorporates accessibility and metadata-extraction functionality
% necessary for future Digital Library endeavors. Numerous ACM and
% SIG-specific \LaTeX\ templates have been examined, and their unique
% features incorporated into this single new template.

% If you are new to publishing with ACM, this document is a valuable
% guide to the process of preparing your work for publication. If you
% have published with ACM before, this document provides insight and
% instruction into more recent changes to the article template.

% The ``\verb|acmart|'' document class can be used to prepare articles
% for any ACM publication --- conference or journal, and for any stage
% of publication, from review to final ``camera-ready'' copy, to the
% author's own version, with {\itshape very} few changes to the source.

\section{Related Works}

\subsection{Object Manipulation in Virtual Reality}
Object manipulation in VR has garnered significant interest among researchers. The most intuitive method is hand-based object manipulation. Beginning in 1996, Poupyrev et al. introduced the GO-GO interaction technique, which facilitates the manipulation of objects both in close proximity and at a distance. This method employs a metaphorical extension of the user's arm combined with a non-linear mapping of hand movements to enhance interactive capabilities~\cite{poupyrev1996go}. Mendes et al. developed a hand-based manipulation technique termed MAiOR, which provides distinct translation and rotation functionalities, thereby achieving the precision of degrees of freedom separation without compromising task completion time~\cite{mendes2017using}. Gloumeau et al. developed PinNPivot, a manipulation technique that utilizes controllers to map hand gestures for virtual hands engaged in manipulation tasks. They compared this technique to other baseline methods, including MAiOR, and found that their approach demonstrated superior performance~\cite{gloumeau2020pinnpivot}. However, hand-based manipulation techniques are widely recognized as more likely to induce arm muscle fatigue, which detrimentally affects the user experience~\cite{jang2017modeling,yu2021gaze}. Consequently, many scholars have explored alternative approaches. Some studies have demonstrated that gaze can effectively support object manipulation. Robert proposed that eye movements could function as an input method for computer interactions~\cite{jacob1990you}. Yu et al. developed a technique for 3D object manipulation that integrates gaze for object selection with manual manipulation for object adjustment~\cite{yu2021gaze}. Furthermore, Bao et al. introduced methods such as Gaze Position, Guided Interaction, and Gaze Beam Guided Interaction, which not only utilize gaze for object selection but also facilitate object movement~\cite{bao2023exploring}. Additionally, several studies have investigated the use of voice commands for object manipulation. For example, Adam S. William et al. conducted an elicitation study on how gestures and speech can be used to manipulate objects in mixed reality environments~\cite{williams2020understanding}. Similarly, Zhou et al. allowed participants to customize their gestures and speech for interactions with multiple objects in another elicitation study~\cite{zhou2022eliciting}. Another approach, proposed by Liu et al., involves using head movements for object manipulation. This method not only reduces user fatigue and motion sickness but also enhances usability and decreases task load~\cite{liu2024object}. Despite these advancements, the primary limitation remains the physical strain and cognitive load associated with the prolonged use of VR systems for object manipulation. These issues are exacerbated in environments requiring complex or repetitive movements, limiting the duration users can comfortably engage with VR. Additionally, while alternative methods like gaze and voice interaction reduce physical strain, they often introduce challenges in terms of precision and control, which can compromise the effectiveness and intuitiveness of interaction in virtual settings. This complexity suggests a need for further research into hybrid interaction techniques that can leverage the strengths of various input methods while minimizing their weaknesses, aiming to enhance the overall user experience in VR applications.

\subsection{Voice-command Interface}
Voice commands are integral to the development of VR interfaces, serving various functions such as navigation~\cite{hombeck2023tell}, design~\cite{morotti2020fostering}, and interactive dialogue~\cite{gobl2021conversational}. The scope of research in voice-enabled VR interfaces is extensive; however, this study narrows its focus to the specific use of voice commands for interacting within VR settings. Notable contributions in this area include Schroeder et al.'s development of a voice-activated system for VR-based alternator maintenance training~\cite{schroeder2017presence}, and Desolda et al.'s implementation of a voice-driven system to aid in 3D modelling ~\cite{desolda2023digital}. Aziz et al. introduced innovative voice-controlled techniques—NoSnap, UserSnap, and AutoSnap—for manipulating graphical object dimensions, demonstrating through user evaluations that these methods, particularly AutoSnap, significantly enhance efficiency and accessibility for users with physical impairments in creative applications~\cite{aziz2022voice}. Additionally, Friedrich et al. introduced an innovative interaction paradigm that merges voice control with hand gesture recognition for intuitive manipulation of CAD models in VR~\cite{friedrich2021combining}. Whitlock et al. investigated the efficacy of various interaction modalities - including voice commands, freehand gestures, and handheld devices - for manipulating objects at different distances in augmented reality~\cite{whitlock2018interacting}. Furthermore, Fernandez et al. developed Hands-Free VR, a natural language voice interface for VR that leverages advanced deep learning models for speech-to-text conversion and sophisticated language models for precise text-to-command translation, demonstrating superior efficiency and user preference compared to traditional VR interfaces~\cite{fernandez2024hands}. Despite these advancements, the integration of voice commands with other control modalities can sometimes create inconsistent user experiences, particularly when switching between interaction types or dealing with complex command structures. These limitations highlight the need for further technical refinement to enhance reliability and user satisfaction in diverse operational settings. 

\subsection{LLM-based Interface}
LLMs have captivated the global research community due to their demonstrated efficacy across various applications. For instance, LLMs have shown potential in enhancing writing skills~\cite{jakesch2023co}, aiding programming tasks for novices~\cite{kazemitabaar2023novices}, and developing question-answering capabilities in children~\cite{abdelghani2024}. These successful implementations often utilize what is known as prompt engineering. As highlighted by Arora et al., effective prompts typically involve question-answering formats that foster open-ended generation. By feeding LLMs with QA examples, they are able to generate stable responses, thereby facilitating their integration into VR environments. However, the deployment of LLMs is highly task-specific, necessitating tailored configurations for different applications. Consequently, designing an efficient LLM interface for VR remains a challenging endeavor.

Several studies have contributed to the development of LLM interfaces in VR, each focusing on different aspects of user interaction and system integration. Wang et al. explored the VirtuWander system, which employs domain-specific LLMs to boost engagement and personalization during virtual museum tours through enhanced multimodal interactions~\cite{wang2024virtuwander}. Wan et al. enhanced human-agent interactions within social virtual environments by developing LLM-based AI agents capable of memory-enhanced, context-aware responses~\cite{wan2024building}. Wei et al. created ChatSim, a system that allows for the editing of photorealistic 3D driving scenes via natural language commands, integrating external digital assets and utilizing a collaborative framework of LLM agents for greater realism and efficiency~\cite{wei2024editable}. Bayat et al. focused on improving the user experience in virtual museums by employing a unified design that includes an Intelligent Virtual Avatar and a Virtual Environment, both powered by an LLM~\cite{bayat2024exploring}. Cheng et al. combined augmented reality, narrative, and LLMs in the "Moon Story" mobile AR application, offering culturally relevant, immersive educational experiences to enhance learning among elementary students~\cite{cheng2024scientific}. Shoa et al. introduced the integration of LLM-based virtual humans, such as a virtual Albert Einstein, into hybrid live events to foster enhanced interaction in multi-user VR settings~\cite{shoa2023sushi}. Finally, John et al. introduced a novel 3D avatar-based assistant that leverages LLM technology for a more engaging and human-like interaction across various applications~\cite{john2024llm}. Despite these innovations, VR interfaces integrating LLMs still face significant challenges, primarily in achieving seamless real-time interactions and maintaining consistent performance across diverse user inputs and environmental contexts. The current limitations also include the need for extensive customization to meet specific application requirements and the complexity involved in managing the interaction between LLM outputs and VR system responses. Further research is needed to address these issues, aiming to create more adaptive, responsive, and user-friendly VR systems that can fully exploit the capabilities of LLMs.

\subsection{Layout Generation}
Traditional methods for generating layouts through optimization heavily depend on prior knowledge of feasible configurations, often derived from procedural modelling or predefined scene graphs~\cite{ma2018language,qi2018human}. These approaches necessitate specialized expertise and exhibit limited adaptability in dynamic settings. Consequently, researchers have explored generative models as a potential solution to these limitations. For instance, Miguel et al. introduced GAUDI, a generative model that facilitates both unconditional and conditional generation of 3D scenes~\cite{bautista2022gaudi}. Handa et al. developed SceneNet, a framework designed to generate annotated 3D scenes, thereby enhancing indoor scene understanding~\cite{handa2016scenenet}. Additionally, Chen et al. proposed SceneDreamer, a novel generative neural hash grid that parameterizes the latent space based on 3D positions and scene semantics~\cite{chen2023scenedreamer}. In the realm of LLMs, new perspectives on text-based layout reasoning have emerged, circumventing the limitations associated with traditional data sets. Feng et al. introduced LayoutGPT, a method that composes in-context visual demonstrations in style sheet language to enhance the visual planning capabilities of LLMs, enabling them to consider layout plans with detailed specifications such as bounding boxes and the orientation of furniture items. Furthermore, Fu et al. developed AnyHome, which utilizes text-based inputs to generate realistic spatial layouts by directing the synthesis of geometry meshes within defined constraints~\cite{wen2023anyhome}. SceneTeller, another innovative approach, employs an LLM-based pipeline to generate high-quality scenes~\cite{ocal2024sceneteller}. These advancements highlight a shift towards more flexible, adaptive layout generation technologies that leverage the power of generative models and language processing.
Our work differs from these efforts in that our goal does not rely on a smart agent to generate a scene. Instead, we aim to develop and design an interface that assists users in VR object manipulation.

\section{VR Mover: A Supportive Natural User Interface for Object Manipulation}
In this section, we discuss the rationale and cognitive theory behind \textit{VR Mover}, and how they effect its interaction design.

\subsection{Spatial Manipulation in the Real World} \label{sec:spatial_manipulation_real_world}
We can visualize how we usually communicate spatial manipulation by imagining that we have movers to help us when moving to a new house. We may do the following:

(1) point at a chair and ask the mover to move it to a specific location by pointing again.

(2) verbally ask the mover to move the dining chairs and the table to a general location by pointing

(3) verbally ask the mover to move a table decoration on top of a dining table. 

(4) gesture in a direction to ask the mover to move an object towards a specific direction.

(5) use the surroundings to describe where to place the object.  

(6) once the mover has finished moving the object, we may fine-tune the final position ourselves.

These are just some of the examples of how we may communicate with others regarding spatial manipulation. Later, we will show how we aim to achieve similar interactions with \textit{VR Mover}. In the next subsection, we will briefly discuss the cognitive theory on some of the expression and communication we used here. 

\subsection{Visual Working Memory}
Whether we want to manipulate an object in VR or attempt to communicate spatial manipulation instructions, it is most likely that we have a mental image and then we manipulate it internally first. This human cognitive process is part of our visual working memory (VWM) \cite{ungerleider1998neural}. VWM is how humans hold, manipulate and interpret information for a variety of everyday tasks \cite{logie2003spatial, baddeley2003working}. How VWM affects our ability to manipulate and handle memory for object location and movement tasks is of particular interest to us. Two governing principles in VWM can provide insights into object manipulation interfaces. 

\myListItem{Chunking:} Humans have a tendency to group objects together to facilitate comprehension and communication. In the context of VWM, we have a process called chunking which will encode information as larger perceptual chunks \cite{miller1956magical, thalmann2019does}. Generally, humans may chunk objects together based on similar colors, locations, and shapes (e.g. chairs of the same set) \cite{doi:10.1111}. Alternatively, when chunking an environment, larger environments may be organized as nested sub-environments \cite{sargent2010chunking}; that is, object and their locations may be chunked together into memories represented as an area (e.g. dining area). 
Thus, in the context of object manipulation, we may infer that people will follow this tendency to group related objects during a manipulation task. Thus, a convenient method to select or refer to a group of objects is a topic worth exploring for object manipulation. 

\myListItem{Coarse-to-fine Processing:}
Another tendency of human nature is to process visual information in a coarse-to-fine manner \cite{sugase1999global}, which is referred to as the so-called coarse-to-fine theory. Recent works have shown that VWM also follows this coarse-to-fine process where the mental image is constructed gradually \cite{gao2013coarse, gao2011coarse}. It should be noted, however, that another work has pointed out that the coarse-to-fine process is not the only cognitive pathway in VWM \cite{ye2020two}. Still, it has been shown that coarse information takes precedence over detailed information \cite{gao2013coarse}. Thus, based on these VWM works, it can be inferred that object manipulation may also involve a coarse-to-fine process.
We should consider a fast coarse placement initially, followed by a fine-tuned adjustment later.

\subsection{Interaction Design}
Here, we discuss how the user can interact with the VR Mover, along with its features and benefits. Some of the interaction design aims to mimic the imagined scenarios discussed in \autoref{sec:spatial_manipulation_real_world}. 

\begin{figure}[h]
    \centering
    \includegraphics[width=0.7\textwidth]{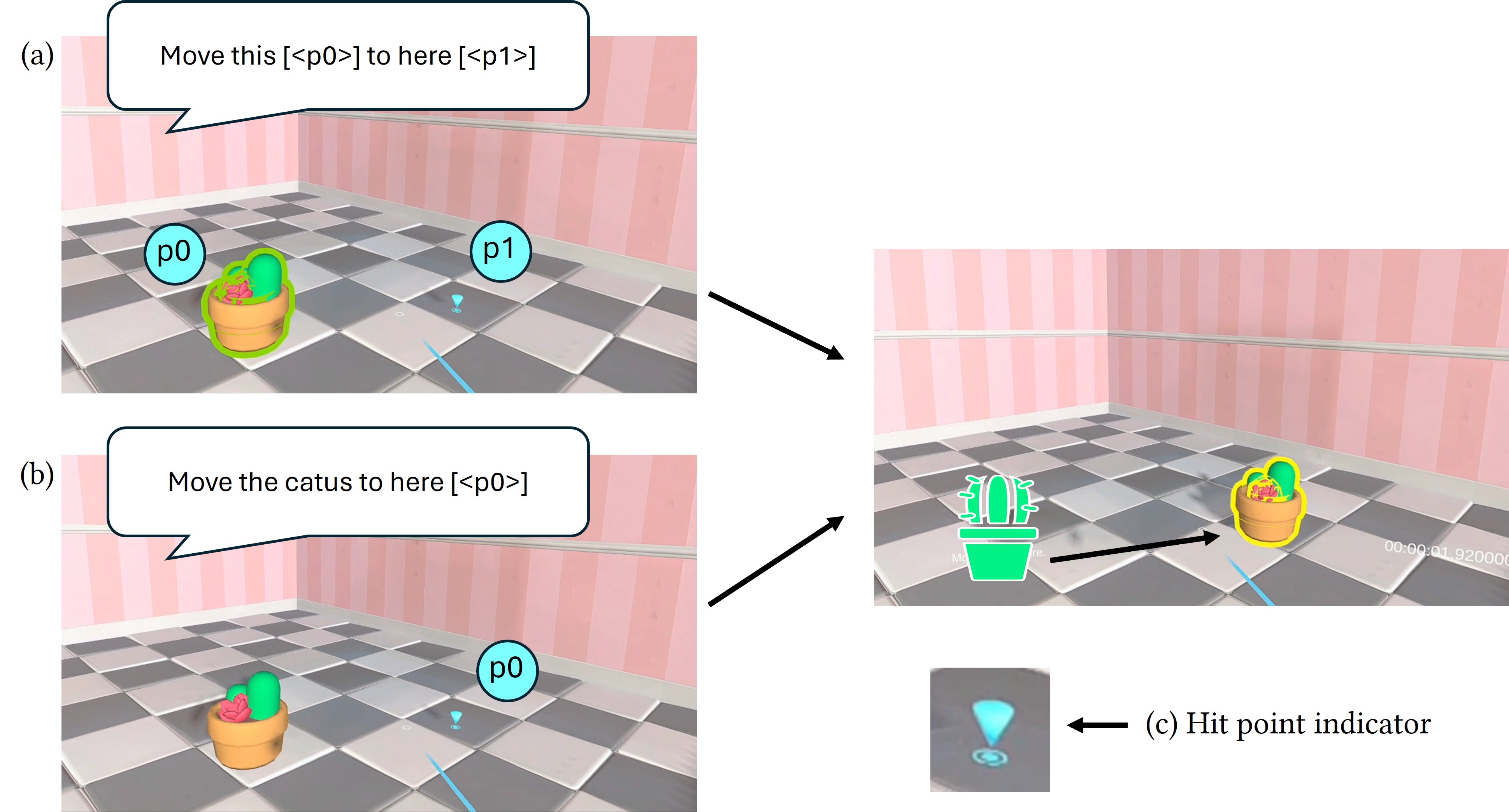}
    \caption{To move an object, (a) the user can first specify an object and its target by pointing. However, as \textit{VR Mover} is aware of what the user is seeing, (b) the user can simply directly use speech to refer to the cactus. Note that, p0 and p1 are the first and second hit points from pointing and (c) a hit point from pointing is visualized.}
    \label{fig:pointing_example}
    \Description{On the left side there are two figures. In the above the user's speech bubble is showing "Move this [<p0>] to here [<p1>]". In the below, the speech bubble is showing "Move the cactus to here." On the right, it is showing the cactus has moved. }
\end{figure}

\myListItem{Pointing:}
In the real world, people naturally use pointing as a method to communicate a point of interest or indicate an object of interest \cite{dalsgaard2021modeling}. In the case of VR Mover, the user can point at an object and then point at a position to either move the object by saying "Move the chair [point] to here [point]" (\autoref{fig:pointing_example}\textit{a}); or to make the object rotate towards that point by saying "Make the chair [point] look at here [point]". It should be noted that our current implementation requires the user to actively quick press A to indicate that they are pointing at something and [point] is indicating when the user has indicated pointing. 

\begin{figure}[h]
    \centering
    \includegraphics[width=0.85\textwidth]{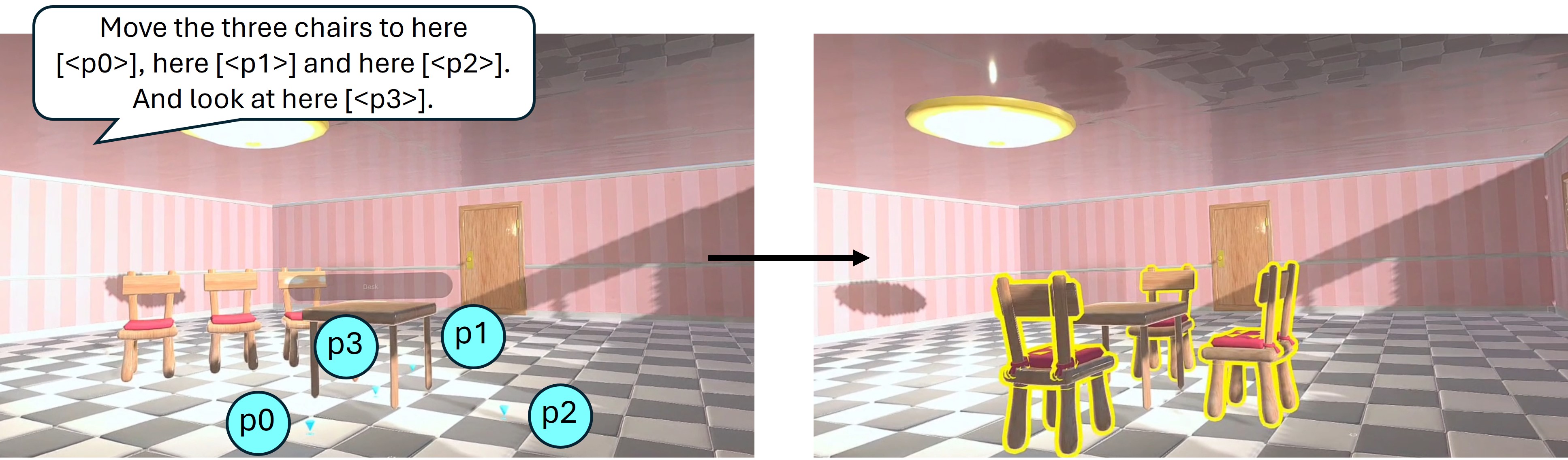}
    \caption{\textit{VR Mover} can handle complex instructions such as the user using asynchronous multi-object manipulation where objects are applied with different manipulation (e.g. different movement) while mixing different manipulation operations (e.g. moving and rotating).}
    \label{fig:multi_obj_manipulation_example}
    \Description{The top is showing that there is a drawn line. And a speech balloon saying "[<d0>] please move the TV along this direction." Then, on the right, the TV is shown to be lifted. The bottom is showing that there is a drawn line over the wall. And on the right, there are four pictures on the wall along the line.}
\end{figure}

\begin{figure}[h]
    \centering
    \includegraphics[width=0.7\textwidth]{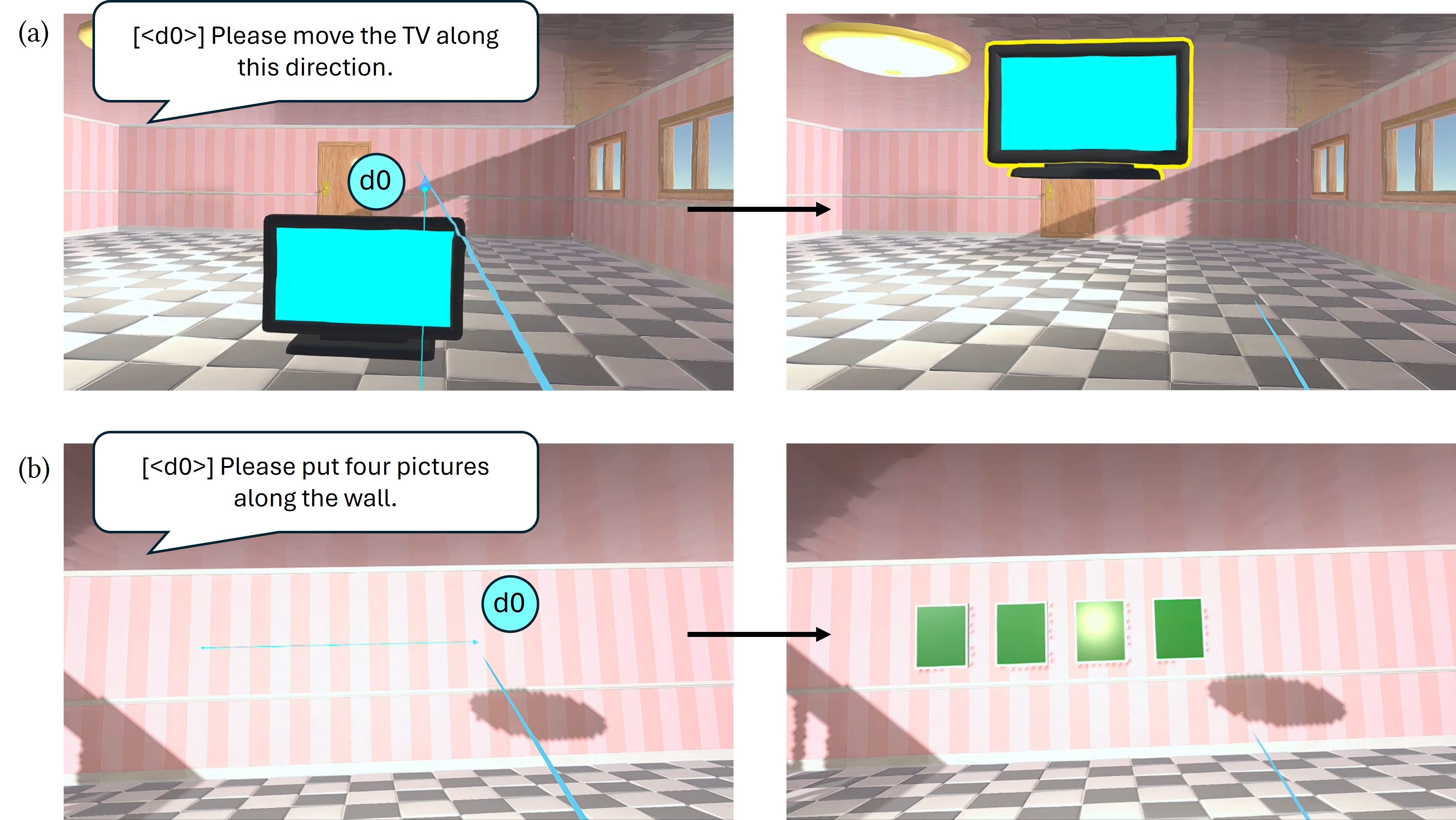}
    \caption{By drawing a line (lining), a user can express different manipulations. Here, we show the user using (a) a line to represent a moving vector, and (b) a line to indicate where the pictures should roughly be placed. \textit{VR Mover} will determine which manipulation is being referred to, based on the user's instructions. }
    \label{fig:lining_example}
    \Description{On the left, three chairs and a table are visible, with a speech bubble saying "Move the three chairs to here [<p0>], here [<p1>] and here [<p2>]. And look at here [<p3>]. On the right, the three chairs are placed at the desired locations, but they have [<p3>]."}
\end{figure}

\begin{figure}[h]
    \centering
    \includegraphics[width=0.7\textwidth]{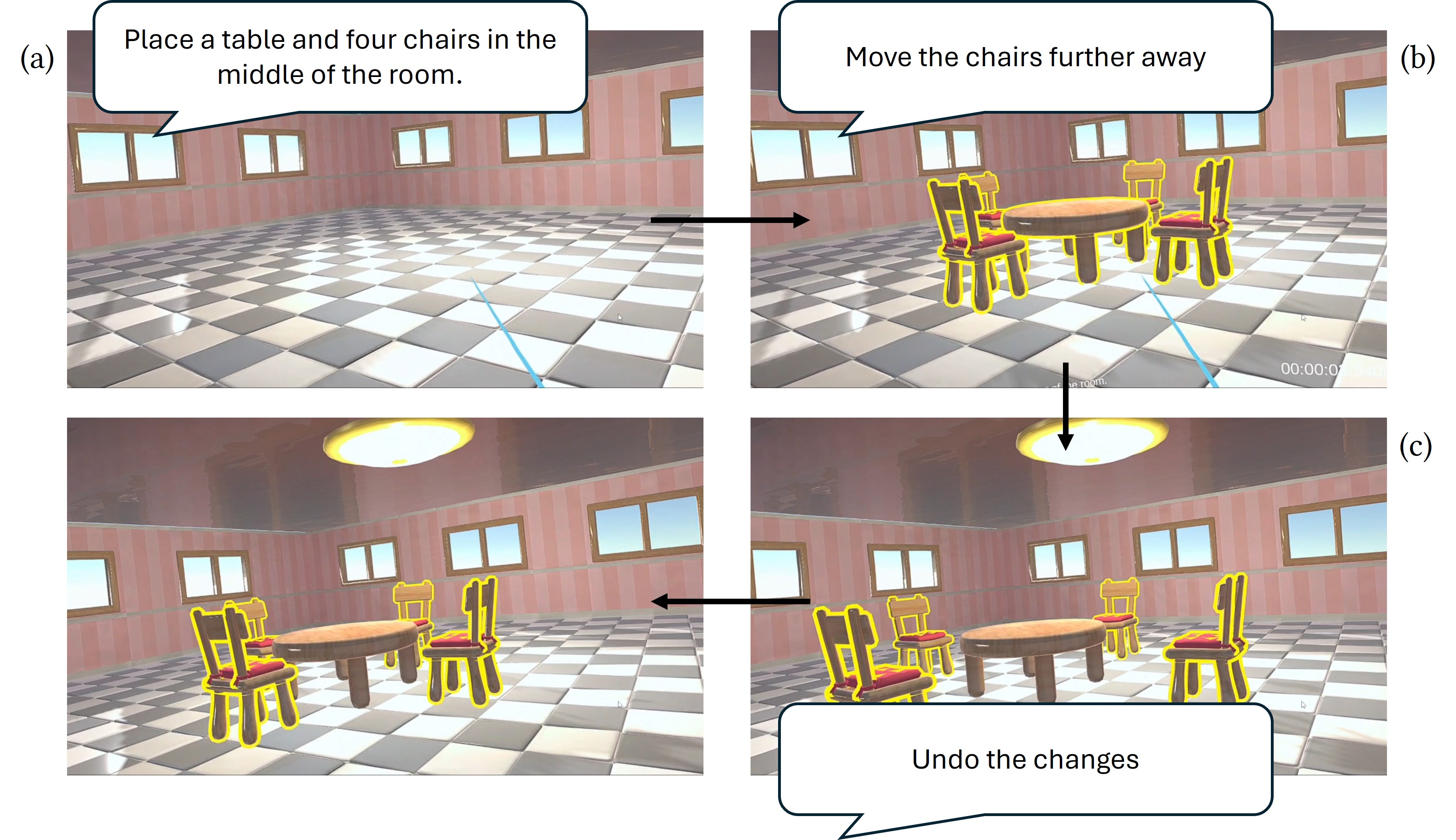}
    \caption{Empowered by LLM, \textit{VR Mover} can demonstrate intelligent responses in some instances. (a) When the user requests four chairs and a table in the middle of the room, \textit{VR Mover} is aware of the environment and able to place the objects in the room's center. Further, it has spatial common sense such that it knows the chairs should be facing the table. (b) As \textit{VR Mover} is aware of the current context as well, when the user is referring to the chairs, it is likely to be the chairs that have just been manipulated. Lastly, (c), although we did not implement an undo function, \textit{VR Mover} is adaptive enough to use the provided APIs to fulfill a user's undo request. }
    \label{fig:context_aware_example}
    \Description{There are four sub-figures. First, there is nothing in the room. The user said ``Place a table and four chairs in the middle of the room". Thus, in the next sub-figure, there are four chairs and a table. At this point, the user said "move the chairs further away". Then, the chairs were moved away. Last, the user said "Undo the changes." }
\end{figure}

\begin{figure}[thp]
    \centering
    \includegraphics[width=\textwidth]{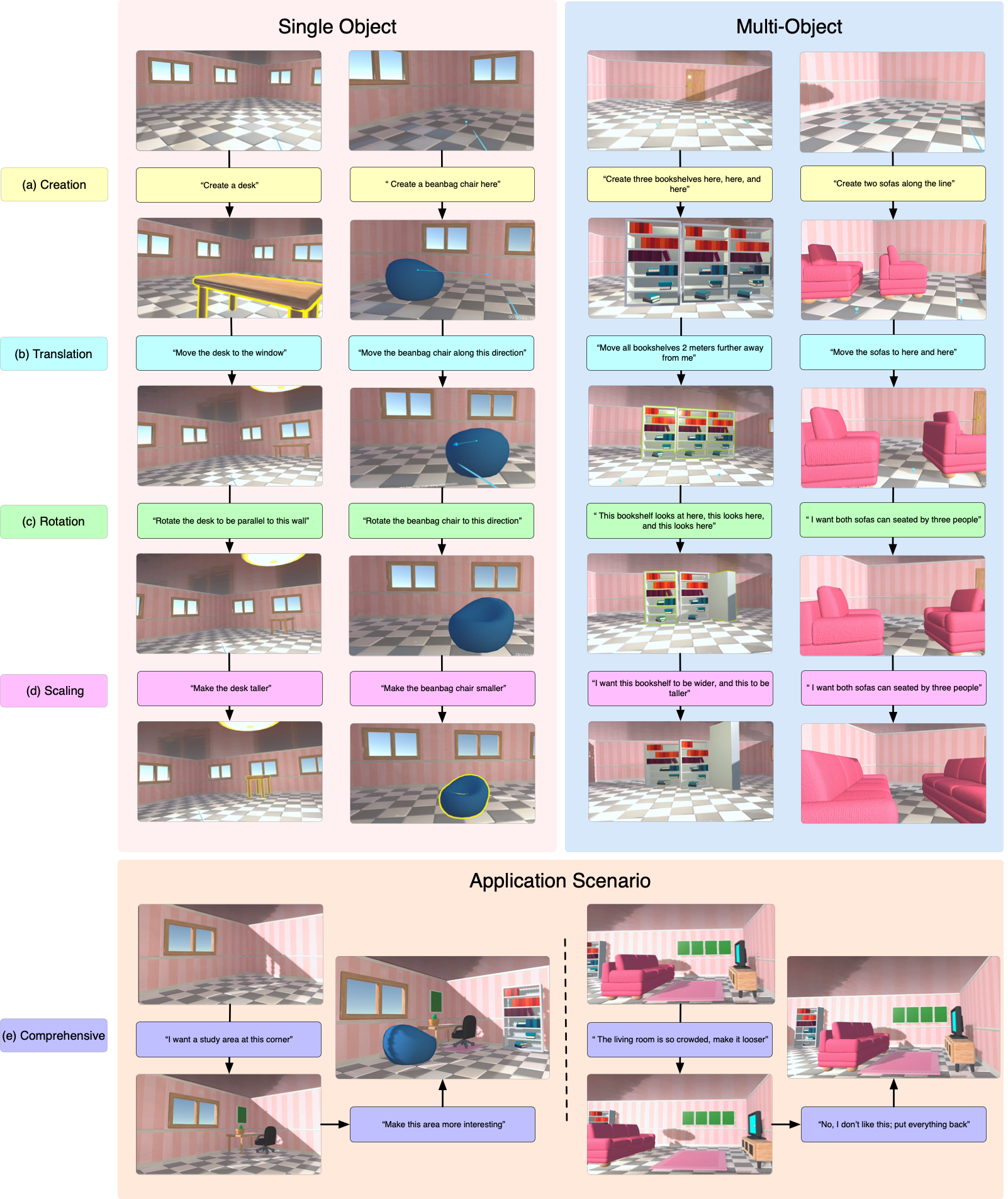}
    \caption{Different interaction methods can be used to engage with \textit{VR Mover}.}
    \label{fig:interact_example}
    \Description{These figures show different methods to interact with the VR Mover. It shows methods for both single-object and multi-object manipulation. At the bottom, there is also an application scenario. For example, the player can ask for a learning area. }
\end{figure}

\myListItem{Lining:}
Aside from pointing, we have different ways to gesturally indicate spatial information. Another method is by drawing a line (lining) to indicate direction or a line-of-interest. For example, the user can say "move the object that way [line]" (\autoref{fig:lining_example}\textit{a}) or "make the object face that direction [line]". Note that similar to pointing, the user needs to press B to draw a line in VR. It can also be used to express an area. The user can say, "I want 4 pictures along the wall here [line]" (\autoref{fig:lining_example}\textit{b}). 

\myListItem{User-centric:}
Similar to a human mover, a \textit{VR Mover} also tries to understand the user's requirements by trying to see the manipulation from the user's perspective. For example, in the case of a chair positioned in front of the user, they can say "move the chair away from me". \autoref{fig:pointing_example}\textit{b} shows an example where the user doesn't need to select an object to manipulate it. 

\myListItem{Asynchronous Multi-object Manipulation:} 
As discussed, humans tend to group things together. With VR Mover, it is possible to quickly refer to multiple objects. They can be explicitly addressed by pointing, or they can be implicitly addressed by saying "move the three chairs to here [point], here [point] and here [point]" (\autoref{fig:multi_obj_manipulation_example}). We referred to this as asynchronous multi-object manipulation because the manipulation applied to each object is different.

\myListItem{Spatial Common Sense:}
As VR Mover is empowered by LLM, it is embedded with a degree of common sense. For example, when asked to place a dining area in a scene, it understands that involves adding chairs and a table. Furthermore, VR Mover will also orientate the chairs to face the table (\autoref{fig:context_aware_example}\textit{a}).

\myListItem{Context-aware:}
As a smart interface, \textit{VR Mover} is aware of the current context. The user may first ask the chair to move to a particular location and then say "move it back". \textit{VR Mover} will still understand what"it" is and where the original location to move it back to is (\autoref{fig:context_aware_example}\textit{b} and 5\textit{c}).

\myListItem{Compound Instruction:}
It is also possible to stack multiple related or unrelated instructions together. For example, we can say "Move the chair to here and then make it face the window"

\myListItem{Manipulation Finetuning:} 
Finally, it should be noted that the intelligent part of \textit{VR Mover} is not designed to complete the entire manipulation autonomously. Just as in real life, where we may ask the mover to move the intended objects to an approximate location, we may later fine-tune the exact placement. Thus, \textit{VR Mover} should be coupled with classical techniques such as gizmos or virtual hands to let the user perform fine-tuning, forming a coarse-to-fine manipulation process.

However, as \textit{VR Mover} is an intelligent interface, there can be many ways to interact with it for object manipulation. We provide more examples of different methods to interact in \autoref{fig:interact_example}.

\section{Methodology of VR Mover}

LLMs have shown promise in spatial arrangements, complex task sequences, and as specialized agents \cite{wang2024chat2layout, wang2024virtuwander, Torre24, bayat2024exploring}. However, many existing models suffer from long response times, ranging from more than 10 seconds to several minutes \cite{10494096, Torre24}, while interactive VR interfaces crucially rely on real-time responses. Additionally, the lack of domain-specific training data, and the resource-extensive fine-tuning of LLMs for the sake of content quality, further complicate the implementation of object manipulation tasks within VR. To this end, we propose an LLM-empowered interface ready for multi-modal object manipulation in virtual space with real-time responses. \textit{VR Mover} follows a user-centric design, it is not only training-free to provide stable responses with around a 2-second delay but can also interpret unstructured instructions into structured object manipulation directives. 

\begin{figure}[b]
    \centering
    \includegraphics[width=\textwidth]{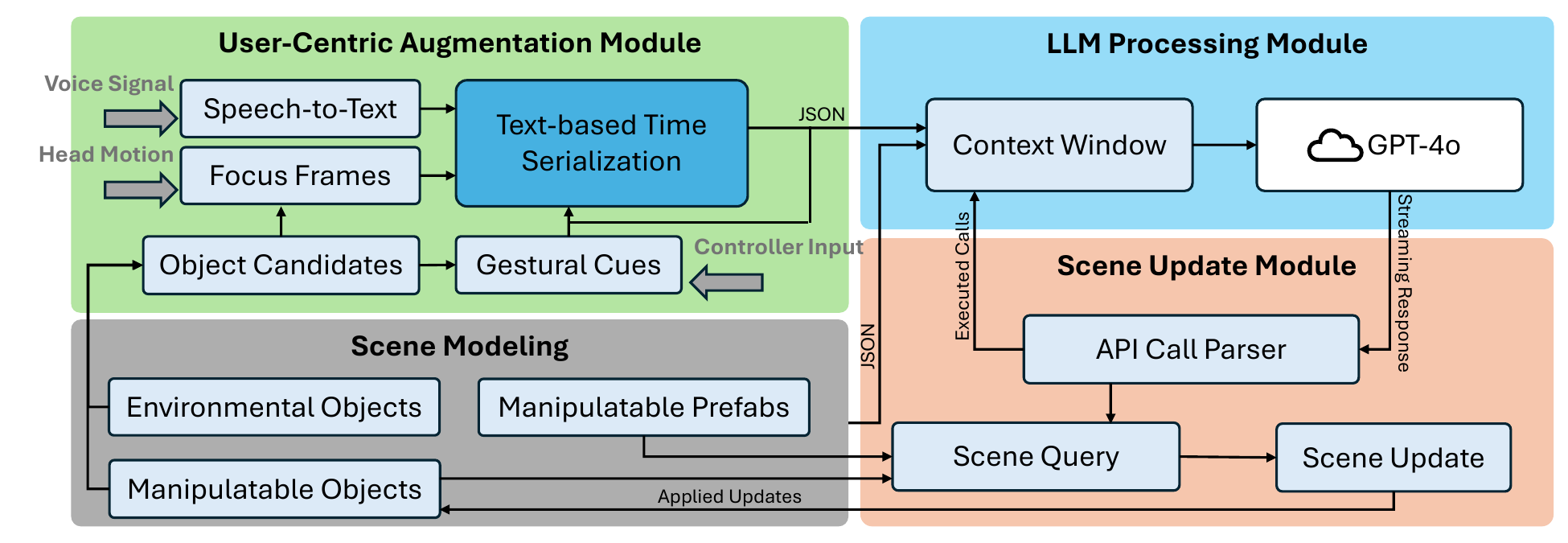}
    \caption{The system overview of VR Mover. The scene modelling component, maintains environmental and manipulatable object data, enhancing the Large Language Model's (LLM) scene understanding. The user-centric augmentation module processes the user's voice, head motion, and controller input, converting them into a JSON format. The LLM processing module, featuring GPT-4o, manages the context and communication with the LLM model. The scene update module interprets LLM responses and updates the virtual environment accordingly.}
    \Description{This is the system overview figure of VR Mover, which consists of four major components: scene modelling (left-bottom), user-centric augmentation module (left-top), LLM processing module (right-top), and scene update module (right-bottom). Taking and arranging the user's inputs including voice, head motion, and controller input, the user-centric augmentation module converts the data into JSON format and sends them to the LLM processing Module. The LLM processing module generates streaming response API calls for the scene update module to parse and execute, where the scene update happens. The scene modelling module converts the 3D scene into text-based expression and can be processed and packed at multiple places in the system.}
    \label{fig:tech-overview}
\end{figure}

Starting from modelling the virtual environment in \autoref{sec:scene-modeling}, \textit{VR Mover} operates in a cycle of three main components: a user-centric augmentation module (\ref{sec:user-centric-augumentation-module}), LLM processing module (\ref{sec:llm-processing-module}), and scene update module (\ref{sec:scene-update-module}). The overview of the interface I/O flow is presented in \autoref{fig:tech-overview}. Scene modelling promotes scene understanding, while the user-centric augmentation module continuously collects motion, speech, and action data from the user. The LLM processing module maintains the text-only prompts with different roles, organizes the context data, and sends requests to the cloud LLM service while streamlining the response. The LLM's response consists of prescribed formatted API calls, which are parsed by the scene update module upon arrival. 

\begin{figure}[b]
    \centering
    \includegraphics[width=\textwidth]{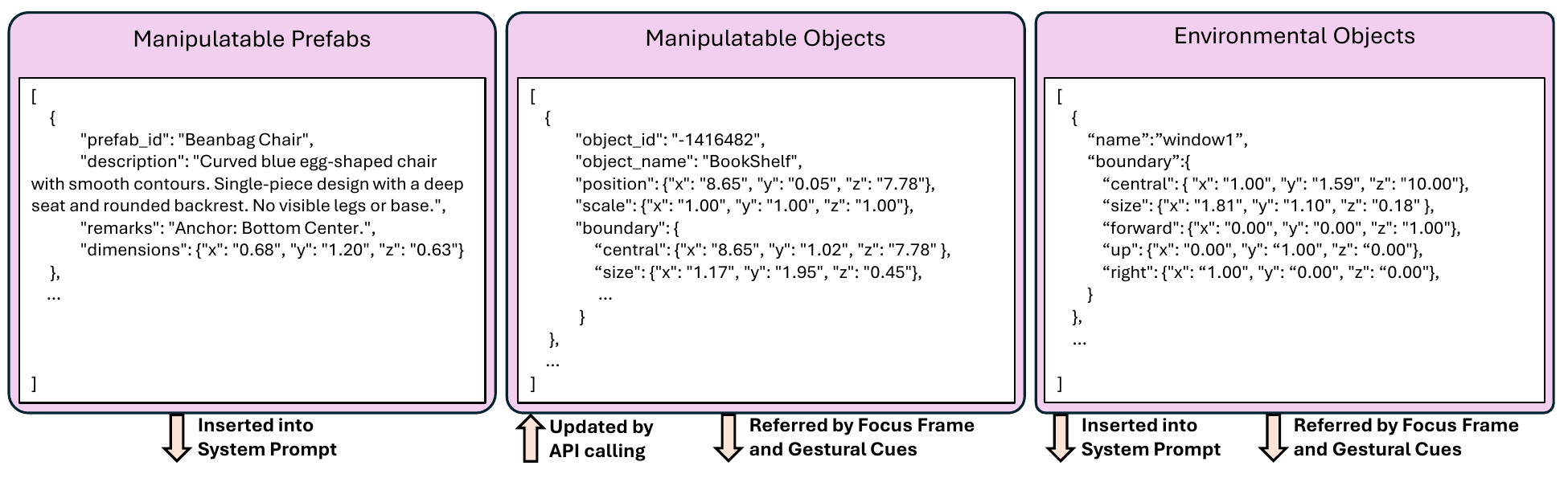}
    \caption{Scene modelling component in VR Mover. Manipulatable prefabs, manipulatable objects, and environmental objects can be expressed in the JSON format, with relevant fields. The arrows indicate where the JSON-ized prefab/objects are referred to or updated from.}
    \Description{Three JSON blocks from left to right, illustrating the JSON format of manipulatable prefabs, manipulatable objects, and environmental objects after being converted from 3D models to text. The manipulatable prefabs contain fields of prefab id, description, remarks, and default dimension for their instances, while the manipulatable objects have the object id, name, position, and scale, as well as the oriented bounding box information. The environmental object only has the name and bounding box information. The manipulatable prefabs and environmental prefabs will be inserted into the system prompt of the LLM context window at the LLM processing module, while the manipulatable object and environmental objects will be referred to in the focus frames and gestural cues in the user-centric augmentation module. The information about manipulatable objects can be updated with LLM API calls.}
    \label{fig:scene-modeling}
\end{figure}

\subsection{Scene Modelling } \label{sec:scene-modeling}
To enable the LLM to generate reasonable object(s) placement proposals, it is necessary to effectively convert 3D spatial information into a text-based format while maintaining the important and expressive features of the scene. We model the scene following a previously proposed taxonomy \cite{10.1145/3635142}, where a scene can be defined as objects placed within a background with spatial and semantic information relations. 
% The virtual environment in the interface is a large empty room for furniture manipulations. 
The scene elements are categorized into environmental objects (static)
% , such as floor, walls, windows), 
and manipulatable objects (dynamic and interactive).
% , mainly furniture items). 
For example, if the virtual environment in the interface is a large empty room for furniture manipulations, the scene elements should be floor, walls, and windows and manipulatable objects should be the furniture. 
Both types are 3D models contributing to the scene's spatial and semantic data. 

The spatial information of the objects is modeled via the oriented bounding boxes (OBBs) \cite{SCHNEIDER2003481}, simplifying each to a cuboid with position, rotation, dimension, and directional vectors. The object names convey the semantic information, while additional descriptions provide clarification for similar types (e.g. sofa vs. couch) or spatial properties (e.g. round table vs. square table), and inter-object relationships (e.g. TV and TV console) of manipulatable objects. Semantic information is defined in the manipulatable prefabs for multiple object instances in the scene sharing the same properties, and the spatial information, which is instance-specific, is maintained separately. Due to the static nature, there are no additional descriptions and prefabs for the environmental objects. All the objects and prefabs are packable into a JSON format and fed to the LLM in different types of prompts. \autoref{fig:scene-modeling} illustrates the JSON field of both types of objects and the manipulatable prefabs, and where they are referenced or updated.

\begin{figure}[b]
    \centering
    \includegraphics[width=\textwidth]{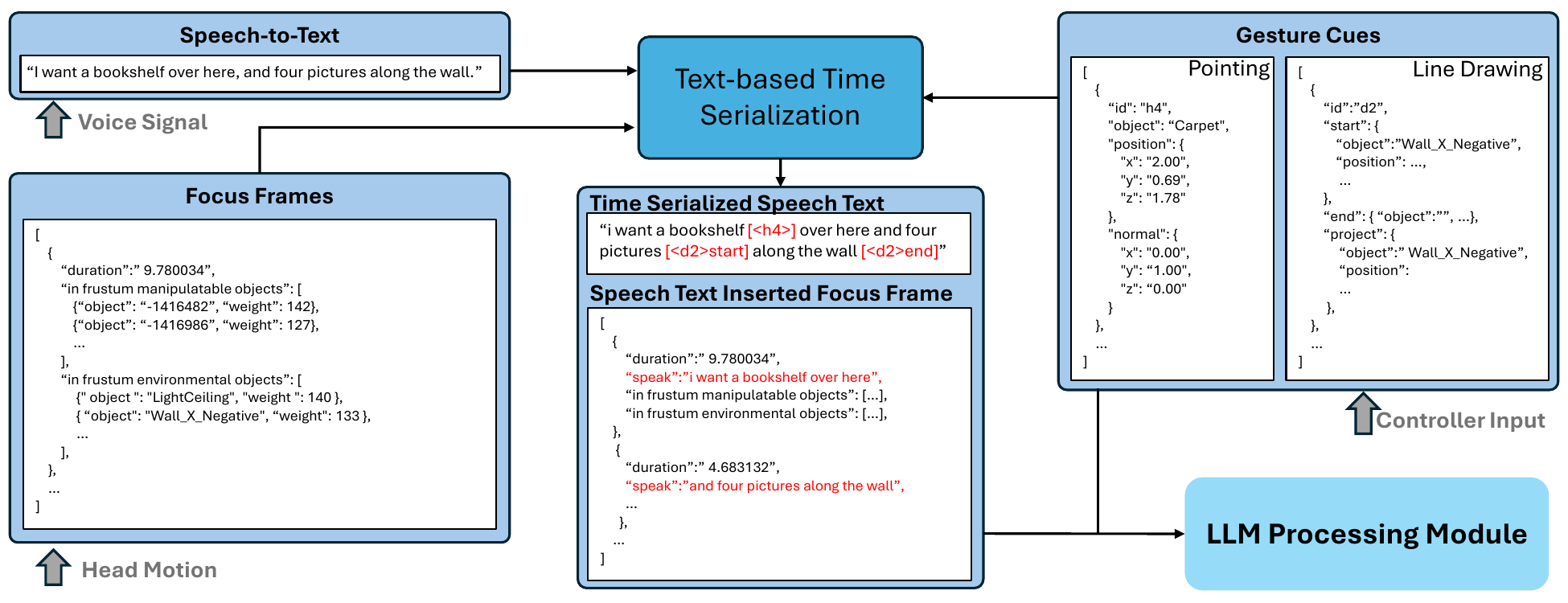}
    \caption{User-centric augmentation module: integrates speech-to-text, focus frames from head motion, and gesture cues from the controller input. Text-based time serialization combines these inputs, producing timestamped data for processing. The example shows the gestural cues being inserted into the speech text, and segmenting the speech text into focus frame groups.}
    \Description{The user-centric augmentation module takes: 1. The voice signal, and transcribes it to text; 2. The head motion data, and collects them as a group of focus frames; 3. The gesture cues from the controller input. The gesture cues can be inserted into the speech text and become a time-serialized speech text, which reflects the timing when the speaking and acting are happening simultaneously. Moreover, the focus frame will be inserted in a new field ``speak'' after text-based serialization as well. This indicates what the user said when looking at this approximate angle. All the actions with augmented elements are passed to the LLM processing module.}
    \label{fig:user-centric-augumentation-module}
\end{figure}

\subsection{User-Centric Augmentation Module}\label{sec:user-centric-augumentation-module}
Maintaining the analogy of communicating with the mover during house moving, the process can be broken down into the following steps: the mover listens to the client's instructions, observes their gestures, and interprets the meaning. It is essential to collect accurate data from the user, organized in a user-centric manner. 
The system primarily prepares data based on: what the user is saying (\ref{subsubsec:speech-recognition}), what the user is looking at (\ref{subsubsec:focus-frames}), where the user has indicated (\ref{subsubsec:processing-gestural-cues}), and what the user is saying or looking at during these actions (\ref{subsubsec:text-based-time-serialization}). This data will be processed and packed into a JSON format (\autoref{fig:user-centric-augumentation-module}) ready to be sent as a part of the user prompt for each request cycle.

\subsubsection{Speech Recognition}\label{subsubsec:speech-recognition}
The speech-to-text (STT) service, provided by Microsoft Azure \cite{azure}, keeps listening to the voice signal captured by the microphone on the VR headset and returns transcriptions with metadata when speech is detected. The return data serves three main purposes: 1. Provide dynamically transcribed text from the user's speech; 2. The timestamp of each spoken word, for future time serializations (\autoref{subsubsec:text-based-time-serialization}); 3. Decide whether the user has stopped talking, and if so, send the packed request to the LLM service to generate a response (\autoref{sec:llm-processing-module}). Furthermore, the transcribed text can be used for interjection filtering. 

\subsubsection{Focus Frames}\label{subsubsec:focus-frames}
%% TODO: Refer to the paper in the theory part that supports: People look at the object when talking about it.
Ambiguities can arise if the LLM has no acknowledgment of what the user is looking at during the talking. Due to headset limitations, we track the user's head motion instead of the gaze. However, continually recording the head motion or objects in the view frustum generates a huge amount of data, and is not straightforward for the LLM to decide the object-of-interest. We define focus frames as groups of continuous viewports over a period in which the user is staring at a small range of objects or positions during speech. The system averages the viewport of the focus frames group while accumulating the objects that appear in the view frustum and ranking them based on the screen distance to the center. When the player's current frame head motion changes beyond the threshold, the current group ends, and the current focus frame moves to the next group. A focus frame group with too short a duration will be filtered out. In each focus frame-group, the higher the ranking is, the more likely the user is looking at the related object during that period. One example can be seen at ~\autoref{fig:user-centric-augumentation-module}.

\subsubsection{Processing Gestural Cues} \label{subsubsec:processing-gestural-cues}
Except for the point position for the pointing, and the start/end points data for the lining, additional information is recorded alongside these actions to strengthen the connection between the action, time, and environment. For the pointing, along with the position of the intersection point between the interaction ray and any of the initial objects in the scene, the surface normal and hit objects are timestamped. For the lining, despite the hit object, normal, and position of the starting point, there are two endpoints. One is the result of applying the hand movement during the drawing to the start point, presenting the visual endpoint of the line. Considering that inaccuracies might arise from manual input, the second end point is the intersection point when the player stops pressing the button, which also contains the position, normal, and hit object. The start time and duration of a drawing line are stored together with other relevant data. However, ambiguities are introduced when there are multiple pointing or drawing lines with no apparent link to the text provided, even if the actions are presented in chronological order. The next sub-subsection solves this problem with timestamps.

\subsubsection{Text-based Time Serialization}\label{subsubsec:text-based-time-serialization}
With the transcribed text, focus frames, and the recorded gestural cues, it can still be difficult to understand the user's intentions. The LLM processes the data over a certain period, so timing is regarded as a crucial factor to reduce the ambiguity between actions or viewports. Aside from the pure transcription of the speech (e.g. ``\texttt{Put the chair here, and I want four pictures along the wall.}''), we also provide a gestural inserted transcribed text (e.g. ``\texttt{put the chair here [<p0>], and [<d0>start] I want four pictures to line [<d0>end] the wall}'' by comparing the timestamps of the gestures with the words, where ``\texttt{p0}'' and ``\texttt{d0}'' are the IDs of the pointing and drawing respectively, and can be referred to the collection of the gestural cues). Moreover, to decipher what was the user talking about when looking at a specific area, the transcription segments are inserted into the focus frames as a JSON field. The changes after the time serialization are highlighted in red in the ~\autoref{fig:user-centric-augumentation-module}.

\subsubsection{Interjection Filtering}
In our pilot study, we found that users tend to express interjections (e.g. ``umm'', ``OK'') when the \textit{VR Mover} has completed the request. The user might be interrupted by a response like ``The request is unclear, please provide specific instructions.'' because the interjections will be transcribed and sent to the LLM without being noted. In addition, this sort of redundant communication stretches the context window and also causes a waste of resources. Aside from asking the user to keep silent, we have also filtered frequently occurring interjections such as ``umm'', ``mmm'', ``OK'', and ``good''. If the transcribed text only consists of interjections, the request will not be sent to the LLM.

\begin{figure}[b]
    \centering
    \includegraphics[width=\textwidth]{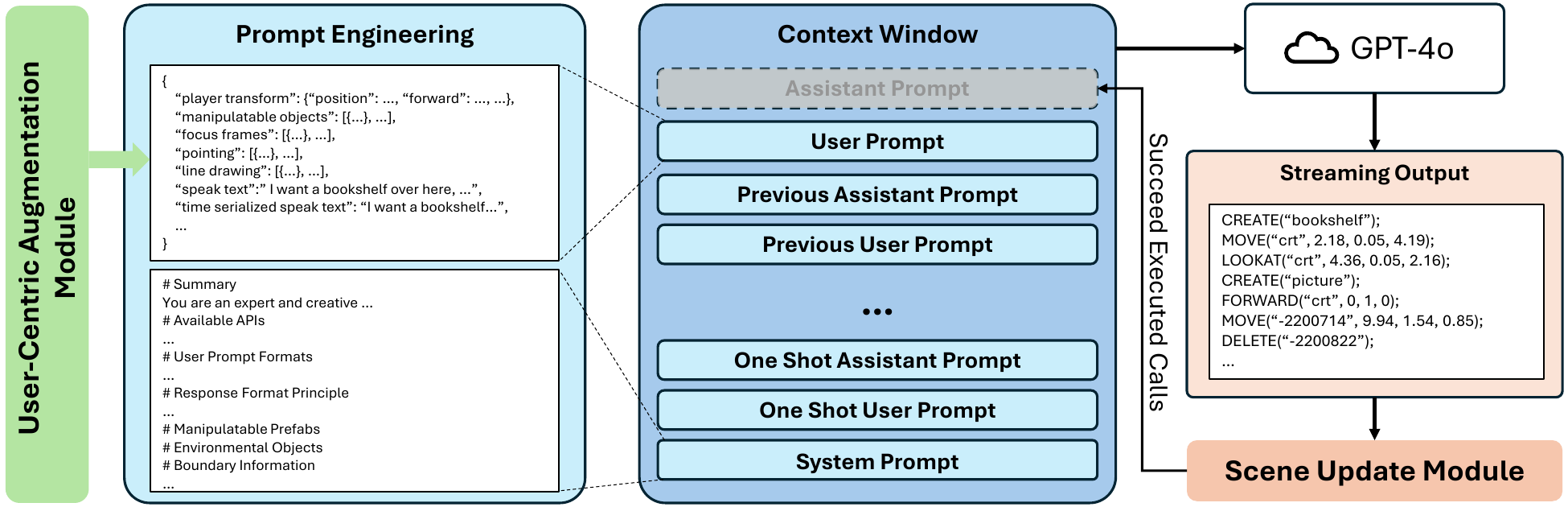}
    \caption{LLM processing module: Prompt engineering combines user inputs with system prompts. The context window manages conversation history. GPT-4o processes and generates streaming output. The output will be passed to the scene update module for processing.}
    \Description{The overview of the LLM processing module. Arranging the JSON from the user-centric augmentation module, the new user prompt is inserted into the context window. The system prompt, along with a pair of shot example user-assistant are always kept at the top of the context window, for defining the roles and controlling the behaviours of the LLM. GPT-4o cloud service starts streaming the API call formatted (specified in the system prompt) after taking the latest context window. The response will be sent to the scene update module line by line for execution, successfully executed calls will be inserted into the latest assistant prompt as record.}
    \label{fig:llm-processing-module}
\end{figure}

\subsection{LLM Processing Module}\label{sec:llm-processing-module}
The overview of this module is shown at \autoref{fig:llm-processing-module}. The background LLM hosted on Microsoft Azure \cite{azure} is responsible for parsing the user's intentions and requests from the user-centric augmentation module (\ref{sec:user-centric-augumentation-module}) in a JSON, and returning generated formatted API calls to perform object manipulation automation. To ensure the LLM behaves properly in a highly dynamic situation and provides logical outputs in a limited time, transforming the user's unstructured requests into a set of well-structured API calls to the structured 3D space management, we instruct it via prompt engineering, manage the context, and use a streamlined response with GPT-4o, a very fast output model. 

\subsubsection{Prompt Engineering}
The three types of role-defined prompts serve different purposes. In our system prompt, except for providing its role and principle, we prepared a list of available APIs for the LLM to generate the response, explained the input format to guide the reasoning, and provided the environmental objects and manipulatable prefabs for it to place and manipulate. Upon each request, the outstanding data collected by the user-centric augmentation module are packed into a JSON format and inserted into the context. The assistant prompt is basically the duplication of the exact response of LLM, filtering out the API calls that fail to run.

However, we found it impossible to generate reproducible \cite{MicrosoftLearn2024} or stable responses for the current settings, even if we locked the seed, and set the temperature and top k variables as 0. The quality of the response varies each time when starting the system. To address this problem, we manually composed a pair of user and assistant prompts (with a detailed explanation before each API call) right after the system prompt, as a one-shot example, in the context to guide the reasoning and fusion. The contents produced by the LLM then tend to fluctuate within a smaller and more acceptable range.

\subsubsection{Context Management}
While \textit{VR Mover} interface conceptually runs for an infinite time and number of communication rounds, typical LLM models' context length is fixed and limited. To avoid context overflow, a sliding window is applied to the context. The system prompt and the example pair of user and assistant prompts stay at the top, and the latest 5 pairs of user-assistant prompts are appended upon each request. Due to the dynamic length of the user prompt, which can be affected by the existing objects in the scene, number of gesture cues, duration of the speech, etc., there is still a risk of overflow, and the LLM server will trim the context. However, according to the practice and experiments, the LLM can redo the scene changes, and construct linkages between consecutive user-assistant prompt pairs.

\subsubsection{Real-time Response Formatting}
Multiple on-the-shelf local and cloud LLM models, series such as Llama \cite{meta_llama}, GPT \cite{openai_chatgpt}, and Claude \cite{anthropic_claude}, are available for performing the tasks through our interfaces. Recent work mainly utilize GPT-4, however, as shown in \cite{wang2024chat2layout}, \cite{10494096}, and \cite{Torre24}, the response time varied from 10 seconds or more to minutes. This is unacceptable for an interactive interface. We chose GPT-4o \cite{openai_gpt4o} hosted by Azure \cite{azure}, for the sake of response speed and generated content quality. Even if GPT-4o accepts image data, and it is possible to perform visual prompting to enhance the placement of objects, image data takes significantly longer to process. Thus, we have chosen to investigate how text-based structures handle spatial data for the sake of quick response time.

Nevertheless, the response time is also related to the formatting of the response data. While several works \cite{wang2024chat2layout, wang2024virtuwander, wei2024editable} prefer to stipulate JSON as the response format, it is not straightforward for extraction if not completed, meaning the system cannot execute the response until the entire JSON is generated and returned. We require the LLM to follow a line-wise API function call in the format of ``\texttt{<function name>(<param1>, <param2>, ...);\textbackslash n}'' (Full list of APIs are shown at \autoref{tab:llm-api-functions}), and configure the web requests to receive a response in a streaming manner. Thus, once the first line of the response is received, it will be sent to the interpreter in the scene update module (\ref{sec:scene-update-module}), and the user can see the scene starts updating immediately. On average, the response delay, counted from the request sent to the LLM to the first line of the API call executed (the system starts to reflect the user's input), is around 2 seconds. With this design, almost all users are satisfied with the response speed.

\begin{figure}[t]
    \centering
    \includegraphics[width=\textwidth]{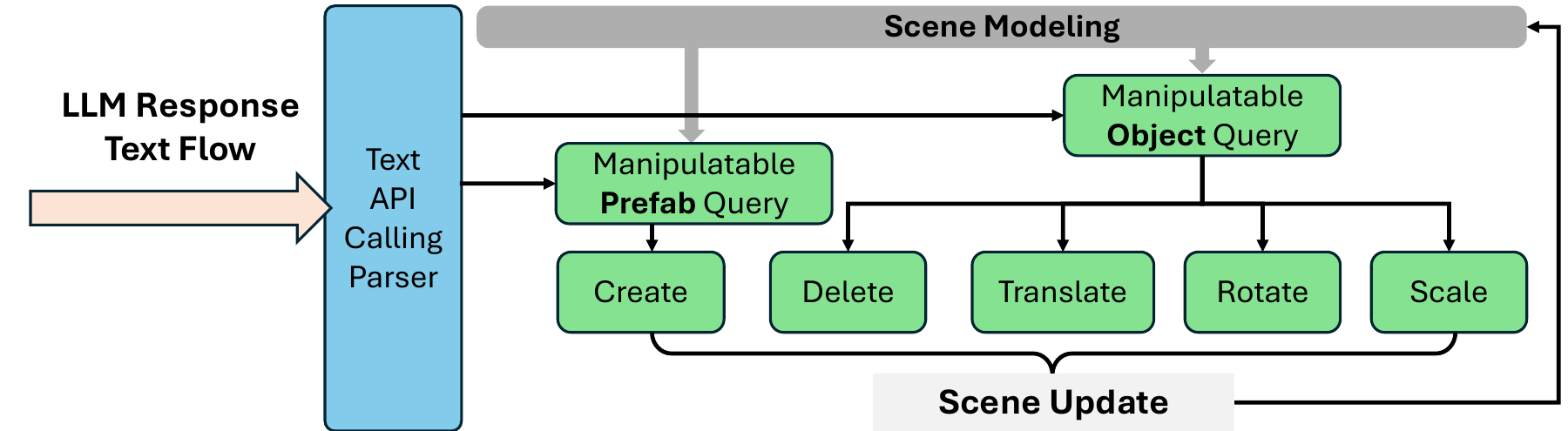}
    \caption{Scene update module: processes LLM responses through text API and parser, queries scene modelling for object data, and executes create, delete, translate, rotate, or scale operations on manipulable objects.}
    \Description{The scene update module is triggered by the LLM response text flow. It parses the API calls line by line, extracting the params and function names to decide which object/prefab to query from the scene, as well as which manipulation type to do, and then performs the manipulation operation, that will affect the scene.}
    \label{fig:scene-update-module}
\end{figure}

\begin{table}[htbp]
\centering
\caption{LLM System Prompt API Functions. The functions provide approaches for the LLM to manipulate objects including create, delete, translate, rotate, and scale object(s). Text-based replies to the user are available when the request is ambiguous or the user asks questions. In addition, a debug utility ``\texttt{EXPLAIN}'' is here for retrieving the inner thinking of the LLM before each function call.}
\Description{This table shows the API calls that are known to VR Mover. On the left there are functions. In the right column, there is the description of the function. }
\label{tab:llm-api-functions}
\begin{tabular}{|p{0.6\textwidth}|p{0.4\textwidth}|}
\hline
\textbf{Function} & \textbf{Description} \\
\hline
CREATE(string prefab\_id); & Create an instance based on the prefab ID \\
\hline
MOVE(string object\_id, float? x = null, float? y = null, float? z = null); & Set the position of an object \\
\hline
FORWARD(string object\_id, float x = 0, float y = 0, float z = 0); & Set the forward direction of the object \\
\hline
LOOKAT(string object\_id, float? x = null, float? y = null, float? z = null); & Set a position for the object to look at \\
\hline
SCALE(string object\_id, float? x = null, float? y = null, float? z = null); & Set the scale of an object \\
\hline
DELETE(string object\_id); & Delete an object by its ID \\
\hline
MESSAGE(string content); & Send a text message \\
\hline
EXPLAIN(string reason); & Send a debug text message \\
\hline
\end{tabular}
\end{table}

\subsection{Scene Update Module}\label{sec:scene-update-module}
This module is in charge of parsing the API calls from the LLM response and updating the scene by either amending the existing object(s) within it or adding new object(s). When a new line is received, the module is triggered and starts to extract the method name and corresponding parameters. The API call will be mapped onto the real runtime functions, corresponding to each type of object manipulation, and invoked asynchronously. A successful API call will affect the virtual scene and update the scene states, in turn influencing the next round of communication. The full flowchart of this module is presented at \autoref{fig:scene-update-module}.

Due to the stochastic nature of the LLM \cite{Torre24}, the LLM is very likely to generate function calls that cannot be parsed and performed (e.g. not existing function, wrong format, or invalid parameters). However, this rarely happens in practice nor has noticeable effects. Therefore, an API call error is not explicitly handled, and the corresponding response entry will not be recorded in the context to mislead future responses.

\section{User Study}
In this section, the setting of the user study is presented. Its result and subsequent discussion will be discussed later. 

\subsection{Experimental Techniques}
In the user study, we will compare the following three techniques. We hope to derive insights from how users used and viewed them differently.

\begin{figure}[h]
    \centering
    \includegraphics[width=\textwidth]{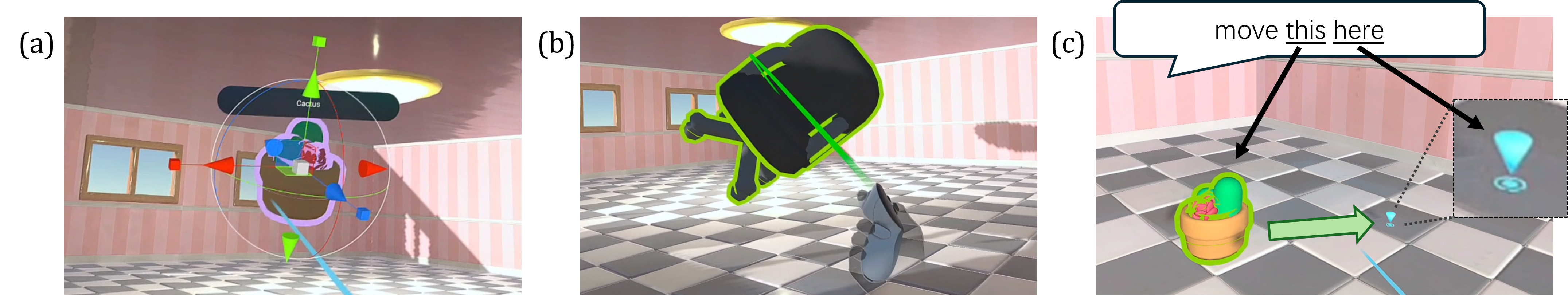}
    \caption{The (a) gizmo and (b) virtual hand are used in all three experimental techniques. The \textit{Voice Command} interface (c) is a LLM-removed variant. It only supported predefined commands such as "move this here". ``this'' has been predetermined to refer to all the selected (highlighted) objects, while ``here'' has also been predetermined to indicate the latest hit point.}
    \label{fig:hand-control-and-voice-command}
    \Description{This figure is showing techniques that are compared with VR Mover. (a) is showing a gizmo. (b) is showing virtual hand. (c) is showing voice command, in which there is an example "move this here. There is an arrow pointing 'this' to the cactus and another arrow  pointing 'here' to the pointing hit point. "}
\end{figure}

\myListItem{Gizmos + Virtual Hand (\textit{Control}):} This is the technique that act as the control. It involves the gizmos (\autoref{fig:hand-control-and-voice-command}\textit{a}) and virtual hand \cite{mendes2019survey} (\autoref{fig:hand-control-and-voice-command}\textit{b}). The gizmos can be interacted with via the interactive ray. For the virtual hand, as it can operate even when the user's hand is not next to the object, it is somewhat similar to the implementation of remote hand \cite{yu2021gaze}. The combination of gizmos and virtual hand is picked as it is believed to be a sufficient representation of object manipulation techniques that are commonly used, see previous research \cite{drey2023investigating, yu2021gaze} and current applications. Multi-object selection to perform synchronized manipulation (e.g. moving all objects in the same way) is possible by first selecting the objects (with quick press A). For brevity, we refer to this interface as the \textit{Control} technique. 

\myListItem{\textit{Voice Command}:} In order to highlight the importance of an LLM that can understand the perspective and unstructured instruction of the user, we have separately developed a voice command variant for the \textit{VR Mover}. Instead of letting an LLM process the user's speech and decide which API to call to complete the object manipulation request, a set of structured voice commands is used to directly map to the same set of APIs. Its implementation is similar to that of a previous voice-driven locomotion technique where the voice command is implemented via regular expression and mapping \cite{hombeck2023tell}. It follows a grammar of <verb, subject, (direction), (unit)>, where the brackets "()" indicate optional input. The most important commands are "move this here" and "rotate this here" in which the user can specify the object(s) to manipulate via selecting and the how to manipulate with pointing. It should be noted that this variant interface only includes pointing, but not aligning as the latter is a method to express ambiguous requests (inferring direction, area, and movement). Thus, it is believed that this is no clear way to implement alignment for Voice Command. 
Regardless, it is not expected to perform well for fine adjustment of object placement. Thus, this \textit{Voice Command} interface also includes the same gizmos and Virtual Hand from the \textit{Control} technique (\autoref{fig:hand-control-and-voice-command}\textit{c}).

\myListItem{\textit{VR Mover}:} The proposed \textit{VR Mover} is an LLM-based interface that can support the user in object manipulation. Similar to \textit{Voice Command}, it is not expected to perform well for fine adjustment of object placement. Thus, it also includes the Gizmos and Virtual Hand. It is expected that the user will use \textit{VR Mover} for a rough or multi-object movement that may or may not involve several combined instructions. The user can then use Gizmos or Virtual Hand for final adjustment to complete the object manipulation.

\subsection{Experimental Tasks}
Similar to a previous VR object manipulation work \cite{yu2021gaze}, our user study includes two tasks. Task 1 is a performance-centric task aimed at evaluating a user's ability to move objects given specific targets. It is further divided into two sub-tasks, Task 1A and 1B. Task 2 is a creative-oriented task that allows the users to freely move objects to embellish the content of a VR room.  

\myListItem{(Task 1A) Single Mid-air Object Manipulation:} The first task involves the user moving an object from source to target. A semi-transparent version of the manipulatable target is used to indicate the goal. The scene involves a chair placed on the ground and a target in mid-air. The distance between the source object and the goal target is measured by the distances between the eight points of a bounding box. Once the average distance is below a threshold, the object is considered to have reached the target. In this sub-task, we aim to evaluate a technique's ability to handle mid-air manipulation. The chair is around one meter tall and both the object and target are three meters away from the user.   

\myListItem{(Task 1B) Multi-object Manipulation:} The second task involves moving several manipulable objects to their targets. Similar to the previous task, semi-transparent targets are used to indicate the goals. 
% the objects are purposely placed in an orderly fashion so that the user can describe the pattern. 
This sub-task differs from the previous one in that it aims to evaluate a technique's ability in handling multi-object manipulation. However, we did not force the user to use the multi-object capability and they may complete the task one object at a time if they find that to be more suitable for them. 

\myListItem{(Task 2) Sandbox Room for Object Placement:} In order to see how users utilize the proposed LLM-supported technique in a more realistic setting, this task provides the users with an empty room. Via a gaze-activated prefab menu in VR, the user can use an assortment of prepared 3D models to populate the room. A soft goal of the task is that the user should try to replicate the mini-room provided via the VR UI panel. However, they are encouraged to test the interaction technique as they see fit. It is believed this gives the user an opportunity to test the object manipulation workflow and provide subjective feedback. As there is no specific end goal for Task 2, each user is simply given 7 minutes to freely manipulate objects in the room. 

\begin{figure}[h]
  \centering
  \includegraphics[width=0.5\linewidth]{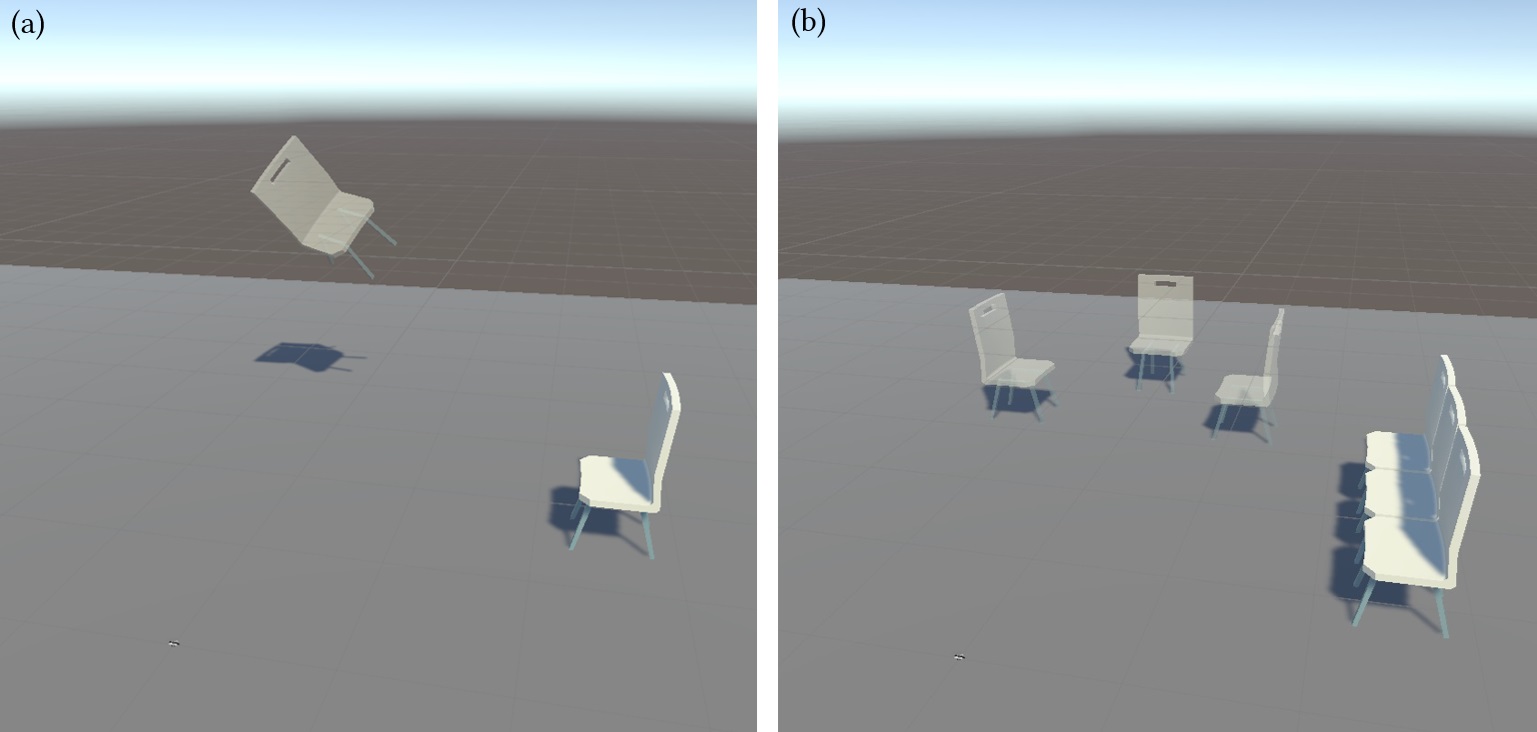}
  \caption{The environment of (a) Task 1A and (b) 1B. The goal of the user is to move the object(s) to the (semi-transparent) target(s).}
  \Description{The figure is presenting the experiment environment of Task 1A and 1B. (a) is showing 1A's, it has a chair on the right and a mid-air target at the center. (b) shows three chairs on the right, and at the center, there are three on-ground goals. }
\end{figure}

\begin{figure}[h]
  \centering
  \includegraphics[width=0.75\linewidth]{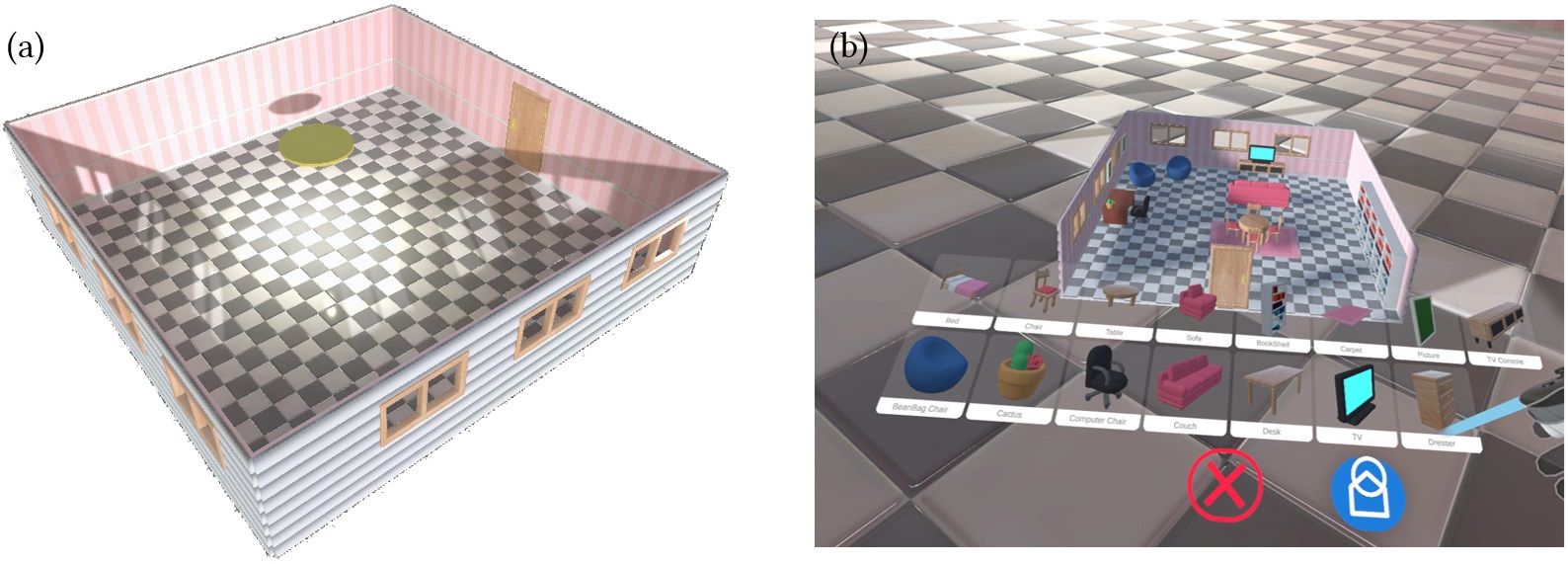}
  \caption{In Task 2, given an (a) empty room the user is instructed to populate it with object according to the (b) mini-room.}
  \Description{The (a) figure is showing an empty room. It is essentially a square room. The (b) figure is showing the user looking down at the UI and see the mini-room when hovering a button. }
\end{figure}

\subsection{Participants and Apparatus}
%Jia Ye, please help to fill this subsection
%Done 
Participants for the user study were recruited through advertisements posted on campus, with all confirming their affiliation as either students or staff of the university. The study involved 24 participants who were randomly assigned to various groups within the experiment. The demographic composition of the study participants included seven males, sixteen females, and one individual of unspecified gender. The mean age of participants was 22.96 years, with a standard deviation of 3.83. Participant ages varied from 18 to 35 years, with a median age of 23 years. Ethical approval for the study was obtained from the institutional review board. 

In the user study, participants experienced VR through a Meta Quest 3 Head-Mounted Display, while experimenters observed user behavior and recorded the content displayed in the headset using a laptop.

\subsection{Measures}
The following measures are used to evaluate the three techniques. Some are objective data collected by the program while some are subjective user feedback. Note that the arrow $\uparrow$ ($\downarrow$) means the higher (lower) the better.

\myListItem{Coarse Manipulation Time ($\downarrow$)} is the time it took for the user to move the object to a "near enough" position (in Task 1). As discussed in a previous paper, the user will spend significant time adjusting the object to fit the target more closely~\cite{yu2021gaze}. This measure provides insights into the time required for a user to move an object from the starting position to a location proximate to the target position. 
An object is considered to have reached to the coarse target when the average distance of the eight points of the bounding box is smaller than 0.3m.

\myListItem{Fine Manipulation Time ($\downarrow$)} is the total duration needed by the user to more closely fit the manipulable object to the target (in Task 1). For fine target, an object is considered to have reached to it when the average bounding box points' distance is smaller than 0.12m. This threshold is chosen based on early testing that most users have significant difficulty reaching the target if it is lower. 

\myListItem{Hand Movement Distance ($\downarrow$)} is the total accumulated hand movement that occurred during Task 1. 

\myListItem{Arm Fatigue ($\downarrow$)} is measured by Borg C10~\cite{borg1982psychophysical,jang2017modeling}, a rating of perceived exertion in the arm.

\myListItem{Usability~($\uparrow$)} is quantified using a modified two-item System Usability Scale (SUS), adapted from the original SUS instrument~\cite{brooke1996sus}. Research by Sauro~\cite{Sauro2018} indicates that items 3 and 8 of this scale can predict the overall SUS score with a 96\% accuracy rate when utilizing a 5-point Likert scale. Consequently, these items were selected to comprise the abbreviated version of the SUS for assessing usability.

\myListItem{Presence ($\uparrow$)} is measured by the Presence Questionnaire (PQ)~\cite{witmer2005factor} which comprises 12 carefully selected items across three pertinent subjects of Presence for our work. These subjects are crucial for evaluating targeting interfaces and include Realism, Quality of Interface, and Self-Evaluation of Performance. Each item is assessed using a 7-point Likert scale.

\myListItem{Workload ($\downarrow$)} is retrieved by the NASA Task Load Index (NASA-TLX) \cite{hart2006nasa}, presented with a 10-points range;

\myListItem{Preference} from users is extracted from a preference rank question where the user will be asked to rank their preference on technique from most favorite ($1^\text{st}$ choice) to least favorite ($3^\text{rd}$ choice).

\myListItem{User Experience ($\uparrow$)} is measured via the short version of the user experience questionnaire (UEQ-S) \cite{schrepp2017design} in a 7-point Likert scale. It captures practical and enjoyable feedback and provides an overall score.

\subsection{Procedure}
For each participant, the following procedure was used: (Step 1) The user fills out a consent form and a basic information questionnaire (e.g. Age). 
(2) Repeat 3a-3e until all three techniques have been tried.
(3a) Based on the random order the user is assigned, the technique's overview is presented to the user. 
(3b) a practice session with an in-VR tutorial to familiarize the user with the interface 
(3c) Complete Task 1
(3d) Complete Task 2
(3e) The user will fill a SUS, Presence Questionnaire, NASA-TLX, and UEQ-S after the completion of a task with a technique. (4) At this point, the user has completed all the trials and will be given the preference ranking to fill out, in addition to an interview session.

\section{Result}
This section reports the results of the user study.
When reporting a measure’s mean (SD) for each technique, the order of reporting is always \textit{Control} group, \textit{Voice Command} and \textit{VR Mover}. 
For each measure, the Shapiro-Wilk test is first conducted to check for data normality. 
If that is the case, typical repeated measures ANOVA and Student's t-tests are used. Otherwise, the Friedman test is performed and the paired test is done with Wilcoxon signed-rank. Bonferroni correction is performed on all post-hoc analyses. Note that we are only interested in comparing \textit{VR Mover} with \textit{Voice Command} and the \textit{Control} group with the post-hoc tests. Effect size is shown with Cohen's d. The threshold for statistical significance was established at a $p - value$ of $0.05$ for all analyses.

\subsection{Manipulation Time (Task 1)}
\begin{figure}[h]
  \centering
  \includegraphics[width=0.75\linewidth]{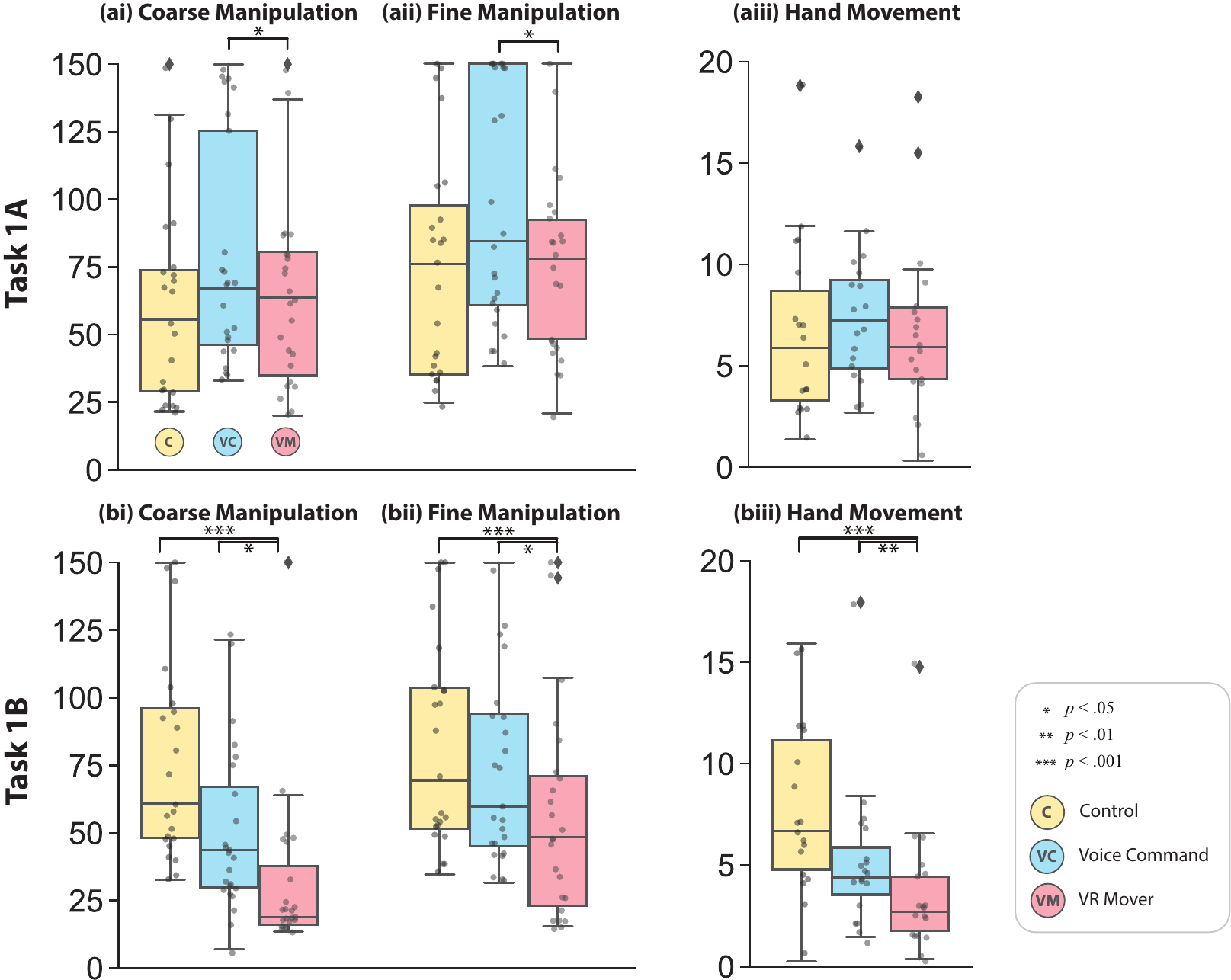}
  \caption{The upper row shows Task 1A's (ai) coarse manipulation time, (aii) fine manipulation time and (aiii) hand movement distance. The lower row shows Task 1B's (bi) coarse manipulation time, (bii) fine manipulation time, and (biii) hand movement distance. For each group, the left is Control, the middle is Voice Command and the right is VR Mover.}
  \Description{The boxplot for Task 1's coarse and fine manipulation time, and the hand movement distance. Mean values and SD are provided in Sec 6.1 and 6.2.}
  \label{fig:studyOneResult}
\end{figure}

For both subtasks of Task 1, we have separately measured the fine manipulation time and coarse manipulation time. For Task 1A, where the user needs to fit a single object to a mid-air target, the mean (SD) of coarse manipulation time (\autoref{fig:studyOneResult}\textit{ai}) is, respectively, $58.142 \text{ (}31.304 \text{)}$, $73.620 \text{ (}32.320 \text{)}$ and $60.765 \text{ (}28.295 \text{)}$, for \textit{Control}, \textit{Voice Command} and \textit{VR Mover}, and the fine manipulation time (\autoref{fig:studyOneResult}\textit{aii}) is $70.682 \text{ (}32.901 \text{)}$, $86.860 \text{ (}30.565 \text{)}$ and $72.11 \text{ (}28.155 \text{)}$, respectively. 
There is a significant main effect for coarse and fine manipulation time. Compared to \textit{Control}, \textit{VR Mover}'s paired test shows no significance for both coarse ($p = 1.00$, $d = 0.0879$) and fine ($p = 1.00$, $d = 0.0728$) manipulation, although the test with \textit{Voice Command} shows a significant difference for coarse ($p = 0.0102$, $d = -0.423$) and fine ($p = 0.0110$, $d = -0.478$) manipulation. This indicates that \textit{VR Mover} does not improve the efficiency for single-object mid-air movement. 

For Task 1B, the user needs to fit multiple objects to on-the-ground targets. The coarse manipulation time (\autoref{fig:studyOneResult}\textit{bi}) for \textit{Control}, \textit{Voice Command} and \textit{VR Mover} is $72.072 \text{ (}28.862 \text{)}$, $50.543 \text{ (}29.532 \text{)}$ and $29.852 \text{ (}23.557 \text{)}$, respectively, and the fine manipulation time (\autoref{fig:studyOneResult}\textit{bii}) is $78.456 \text{ (}30.067 \text{)}$, $69.988 \text{ (}29.652 \text{)}$ and $52.636 \text{ (}32.361 \text{)}$. There are main effects for both coarse ($F(2,23)=21.583$, $p<0.001$, $\eta_p^2=0.450$) and fine ($F(2,23)=8.769$, $p=0.0125$, $\eta_p^2=0.183$) manipulation time. For coarse manipulation time, \textit{VR Mover}'s paired tests show significant difference with both \textit{Voice Command} ($p = 0.0130$, $d = -0.774$) and \textit{Control} technique ($p < 0.001$, $d = -1.603$).
Similarly, for fine manipulation time, \textit{VR Mover} shows significance compared with both \textit{Voice Command} ($p = 0.0342$, $d = -0.556$) and \textit{Control} ($p < 0.005$, $d = -0.827$). These results indicate that \textit{VR Mover} improves the efficiency when multi-object movement is involved. 

\subsection{Hand Movement Distance (Task 1)}
Aside from the completion time as discussed, we have also measured the total accumulated hand movement in Task 1A and 1B. Note that the hand movement is recorded before the user has reached the fine target goal. 
For Task 1A, the mean of hand movement distances (\autoref{fig:studyOneResult}\textit{aiii}) for \textit{Control}, \textit{Voice Command} and \textit{VR Mover} are $6.400 \text{ (}3.253 \text{)}$, $6.367 \text{ (}2.383 \text{)}$ and $6.533 \text{ (}3.305 \text{)}$, respectively. There is no main effect when different techniques are used ($F(2,23)=1.083$, $p=0.582$, $\eta_p^2=0.0226$). Thus, there is no need to show the paired tests. 

On the other hand, for Task 1B, the hand movement distance is more differentiating (\autoref{fig:studyOneResult}\textit{biii}). The mean movement distances are $7.503 \text{ (}3.405 \text{)}$, $5.148 \text{ (}3.101 \text{)}$ and $3.550 \text{ (}2.740 \text{)}$ for \textit{Control}, \textit{Voice Command} and \textit{VR Mover}, respectively, and there is a main effect for technique ($F(2,23)=22.750$, $p<0.001$, $\eta_p^2=0.474$). Compared to \textit{Voice Command} and Control, \textit{VR Mover} shows statistical significance. Further, paired tests show that the hand movement distance in \textit{VR Mover} is significantly different to that of \textit{Voice Command} ($p < 0.005$, $d = -0.546$) and \textit{Control} ($p < 0.001$, $d = -1.279$). Similar to the previous result on coarse and fine manipulation time, the result here indicates that \textit{VR Mover} has a more pronounced effect when multiple objects are involved.

\subsection{Arm Fatigue}
\begin{figure}[h]
  \centering
  \includegraphics[width=0.6\linewidth]{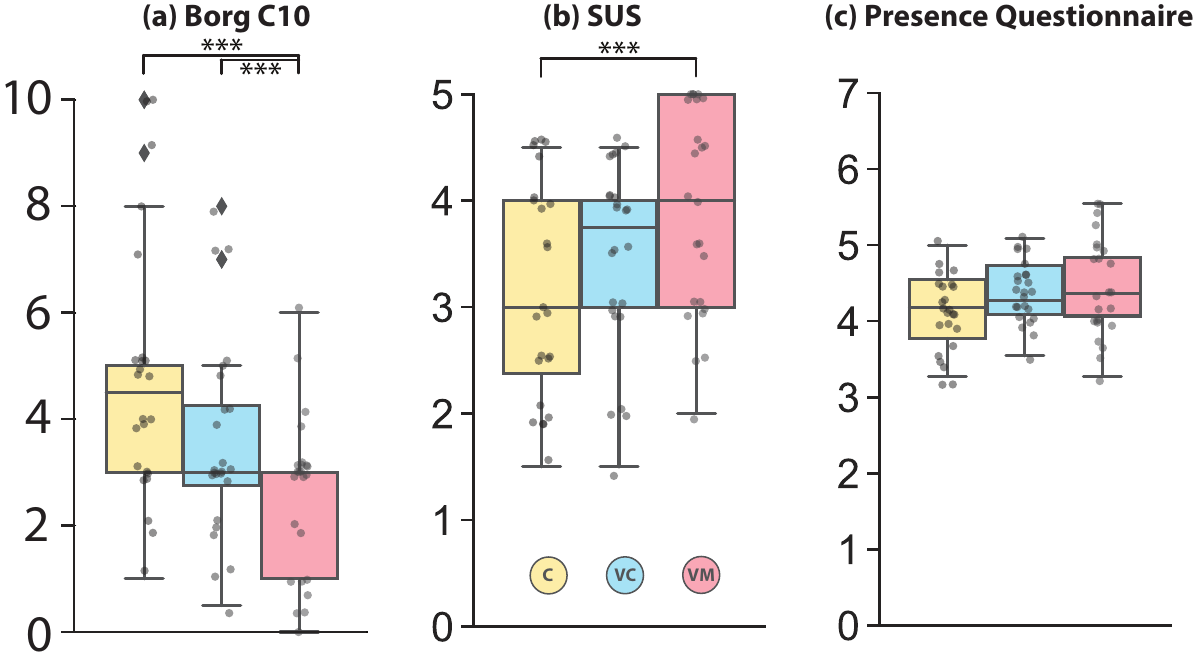}
  \caption{The result from (a) the borg C10, (b) SUS and (c) presence. }
  \Description{The boxplot for borg C10, SUS and presence. Their means and SDs are presented in Sec 6.3, 6.6 and 6.7, respectively.}
  \label{fig:borg_sus_presence}
\end{figure}

Arm fatigue is measured by Borg C10. The average Borg C10s scores (\autoref{fig:borg_sus_presence}\textit{a}) for \textit{Control}, \textit{Voice Command} and \textit{VR Mover} are $4.792 \text{ (}2.380 \text{)}$, $3.521 \text{ (}1.857 \text{)}$ and $2.521 \text{ (}1.461 \text{)}$, respectively. There is a significant main effect for technique ($F(2,23)=20.840$, $p<0.001$, $\eta_p^2=0.434$) and similarly, paired test that compared \textit{VR Mover} with the other two techniques, \textit{Voice Command} ($p = 0.0197$, $d = -0.599$) and \textit{Control} also show significance ($p < 0.001$, $d = -1.150$). 

\subsection{Workload}
\begin{figure}[h]
  \centering
  \includegraphics[width=0.8\linewidth]{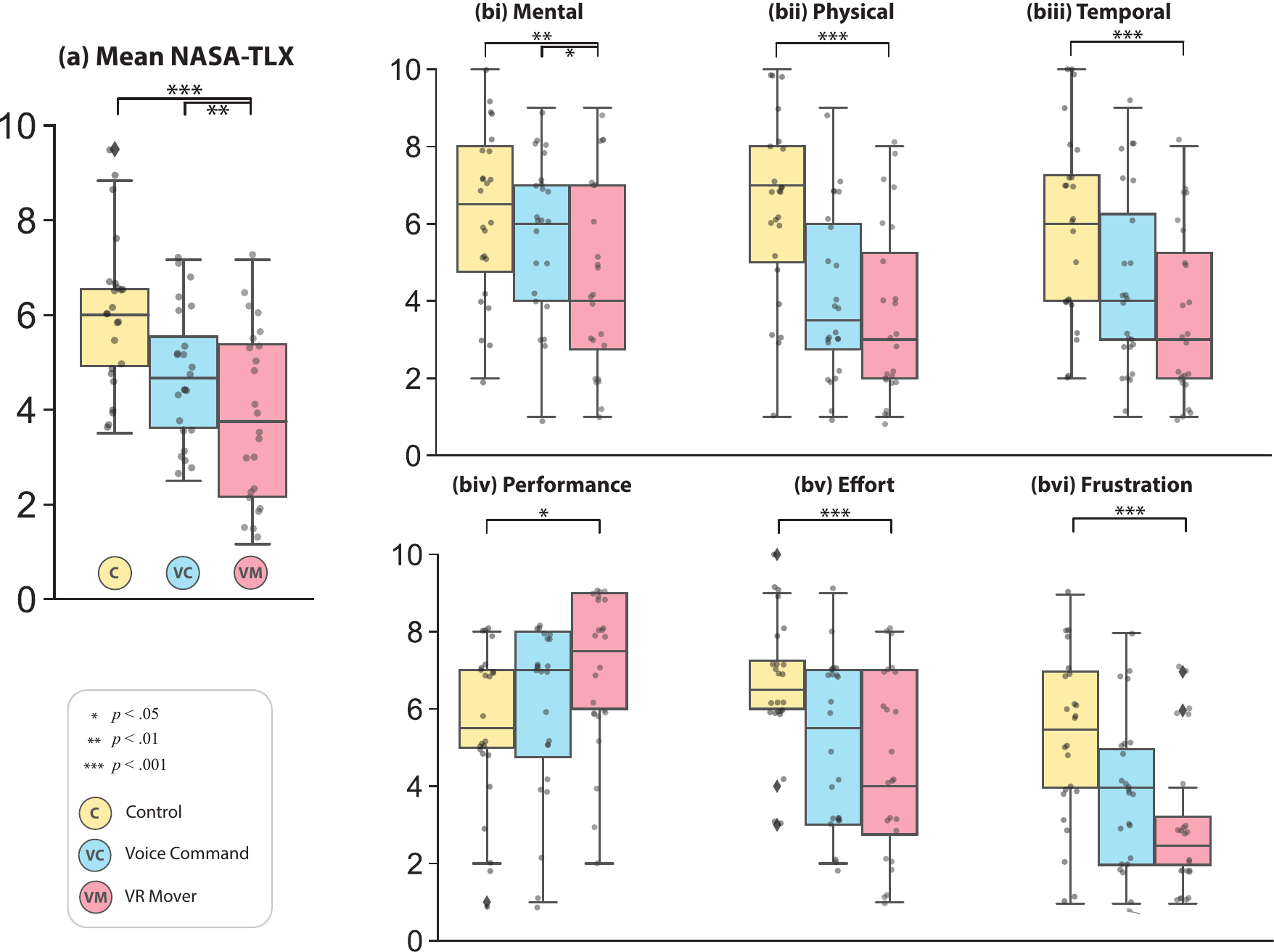}
  \caption{The (a) overall mean score of NASA-TLX and its subscales, (bi) mental demand, (bii) physical demand, (biii) temporal demand (biv) performance, (bv) effort, and (bvi) frustration.}
  \Description{The boxplots of NASA-TLX's mean and the subscales. The means and SDs are presented in Sec. 6.4.}
  \label{fig:tlx}
\end{figure}

The mean NASA-TLX scores (\autoref{fig:tlx}\textit{a}), for \textit{Control} group, \textit{Voice Command} and \textit{VR Mover}, are $5.965 \text{ (}1.588 \text{)}$, $4.729 \text{ (}1.374 \text{)}$ and $3.882 \text{ (}1.797 \text{)}$, respectively. There is a significant main effect for the technique ($F(2,23)=17.725$, $p<0.005$, $\eta_p^2=0.198$). VR Mover's comparison with both \textit{Voice Command} ($p = 0.0409$, $d = -0.530$) and \textit{Control} group ($p < 0.001$, $d = -1.228$) both shows statistical significance. 

To further investigate the constituent components of the users' workload, the data for each subscale is also presented. For mental demand (\autoref{fig:tlx}\textit{bi}), the scores for Control, \textit{Voice Command} and \textit{VR Mover} are $6.208 \text{ (}2.121 \text{)}$, $5.750 \text{ (}1.984 \text{)}$ and $4.542 \text{ (}2.398 \text{)}$, respectively, and it has a main effect ($F(2,23)=7.238$, $p=0.0268$, $\eta_p^2=0.151$). \textit{VR Mover} shows significance compared to both \textit{Voice Command} ($p = 0.0478$, $d = -0.549$) and \textit{Control} ($p < 0.01$, $d = -0.736$) in the paired tests.
The scores 
for physical demand (\autoref{fig:tlx}\textit{bii}) are $6.375 \text{ (}2.324 \text{)}$, $4.125 \text{ (}2.147 \text{)}$ and $3.583 \text{ (}2.253 \text{)}$ 
($F(2,23)=18.025$, $p<0.001$, $\eta_p^2=0.376$),
for temporal demand (\autoref{fig:tlx}\textit{biii}) are $5.958 \text{ (}2.423 \text{)}$, $4.417 \text{ (}2.272 \text{)}$ and $3.583 \text{ (}2.197 \text{)}$ 
($F(2,23)18.024$, $p<0.001$, $\eta_p^2=0.375$), 
for performance (\autoref{fig:tlx}\textit{biv}) are $5.583 \text{ (}1.998 \text{)}$, $5.917 \text{ (}2.197 \text{)}$ and $6.958 \text{ (}1.968 \text{)}$ 
($F(2,23)=7.932$, $p=0.0190$, $\eta_p^2=0.165$), 
for effort (\autoref{fig:tlx}\textit{bv}) are $6.625 \text{ (}1.703 \text{)}$, $5.083 \text{ (}2.159 \text{)}$ and $4.500 \text{ (}2.363 \text{)}$ 
($F(2,23)=10.659$, $p<0.005$, $\eta_p^2=0.222$) and 
for frustration (\autoref{fig:tlx}\textit{bvi}) are $5.208 \text{ (}2.160 \text{)}$, $3.917 \text{ (}1.913 \text{)}$ and $3.042 \text{ (}1.903 \text{)}$ 
($F(2,23)22.747$, $p<0.001$, $\eta_p^2=0.474$), 
and they all have a main effect on technique.
Paired tests also show \textit{VR Mover} has significance when compared with \textit{Control} for physical demand ($p < 0.001$, $d = -1.220$), temporal demand ($p < 0.001$, $d = -1.027$), performance ($p = 0.0191$, $d = 0.693$), effort ($p < 0.005$, $d = -1.032$) and frustration ($p < 0.001$, $d = -1.064W$). However, when comparing with \textit{Voice Command} in physical demand ($p = 0.457$, $d = -0.246$), temporal demand ($p = 0.0791$, $d = -0.373$), performance ($p = 0.168$, $d = 0.499$), effort ($p = 0.385 $, $d = -0.255$) and frustration ($p = 0.0542$, $d = -0.459$), there is no significance.
Overall, the considerably lower NASA-TLX scores of \textit{VR Mover} is a strong indication that it can reduce workload. 

\subsection{User Experience}
\begin{figure}[h]
  \centering
  \includegraphics[width=0.65\linewidth]{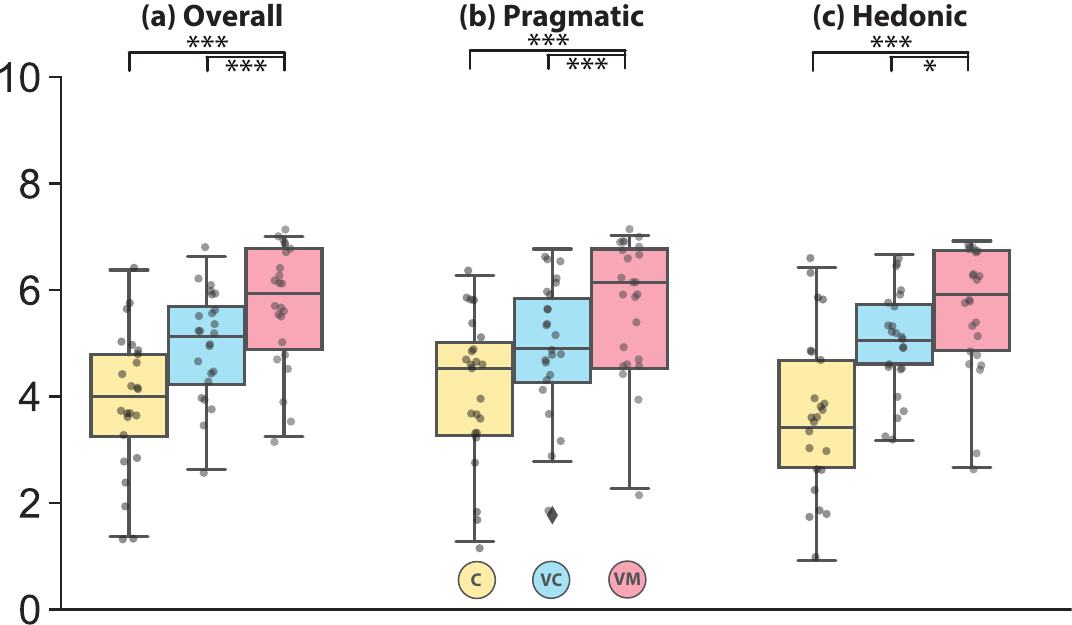}
  \caption{The (a) overall, (b) pragmatic, and (c) hedonic user experience result from UEQ-S.}
  \Description{The boxplot for UEQ-S and its meta-measures. The means and SDs for the overall score and meta-measures are presented in Sec 6.5.}
  \label{fig:ueq}
\end{figure}

The overall UEQ-S scores (\autoref{fig:ueq}\textit{a}) for \textit{Control} group, \textit{Voice Command} and \textit{VR Mover}, are $3.922 \text{ (}1.273 \text{)}$, $4.990 \text{ (}0.953 \text{)}$ and $5.729 \text{ (}1.127 \text{)}$, respectively. There is a significant main effect for the technique ($F(2,23)=26.053$, $p<0.005$, $\eta_p^2=0.278$). The paired tests show VR Mover's comparisons with both \textit{Voice Command} ($p < 0.005$, $d = 0.699$) and \textit{Control} group ($p < 0.001$, $d = 1.494$) have statistical significance.

UEQ-S can be further separated into two meta-measures, practical and hedonic. The former indicates efficiency, perspicuity, and dependability while the latter is for indicating stimulation and novelty. For pragmatic quality (\autoref{fig:ueq}\textit{b}), the scores are $4.125 \text{ (}1.342 \text{)}$, $4.885 \text{ (}1.235 \text{)}$ and $5.719 \text{ (}1.249 \text{)}$, respectively. Similar to the overall score, there is a main effect ($F(2,23)=18.065$, $p<0.001$, $\eta_p^2=0.376$) and the paired test shows significance when comparing \textit{VR Mover} with \textit{Voice Command} ($p < 0.005$, $d = 0.671$) and \textit{Control} ($p < 0.005$, $d = 1.229$).
For hedonic quality, the respective scores (\autoref{fig:ueq}\textit{c}) are $3.719 \text{ (}1.485 \text{)}$, $5.094 \text{ (}0.935 \text{)}$ and $5.719 \text{ (}1.191 \text{)}$, and it also has main effect ($F(2,23)=31.303$, $p<0.005$, $\eta_p^2=0.652$). The paired tests that compared \textit{VR Mover} with \textit{Voice Command} ($p = 0.0183$, $d = 0.584$) and \textit{Control} ($p < 0.005$, $d = 1.486$) both show significance. 

\subsection{Usability}
The ease-of-use is measure by the SUS; the scores (\autoref{fig:borg_sus_presence}\textit{b}), for \textit{Control} group, \textit{Voice Command} and \textit{VR Mover}, are $3.167 \text{ (}0.986 \text{)}$, $3.479 \text{ (}0.884 \text{)}$ and $3.896 \text{ (}0.957 \text{)}$, respectively. There is a significant main effect for the technique ($F(2,23)=17.289$, $p<0.005$, $\eta_p^2=0.360$). \textit{VR Mover} only shows significance when compared with \textit{Control} ($p < 0.005$, $d = 0.750$) group. There is no significant different with \textit{Voice Command} ($p = 0.884$, $d = 0.452$).

\subsection{Presence}
Measure by the PQ, the average scores for presence (\autoref{fig:borg_sus_presence}\textit{c}) are $4.121 \text{ (}0.496 \text{)}$, $4.398 \text{ (}0.379 \text{)}$ and $4.447 \text{ (}0.624 \text{)}$, for Virtual Assistant, \textit{Voice Command} and \textit{Control} group, respectively. There is a significant main effect for the technique ($F(2,23)=4.748$, $p=0.0134$, $\eta_p^2=0.0459$). 
However, there is no significance when comparing \textit{VR Mover} with either \textit{Voice Command} ($p = 0.598$, $d = 0.0954$) and \textit{Control} ($p = 0.0651$, $d = 0.578$) in paired test. Still, it is believed that given the existence of the main effect, the paired test comparing \textit{VR Mover} shows marginal significance and that the effect size is considerable, further investigation should be considered.

\subsection{Ranking}
\begin{figure}[h]
  \centering
  \includegraphics[width=0.5\linewidth]{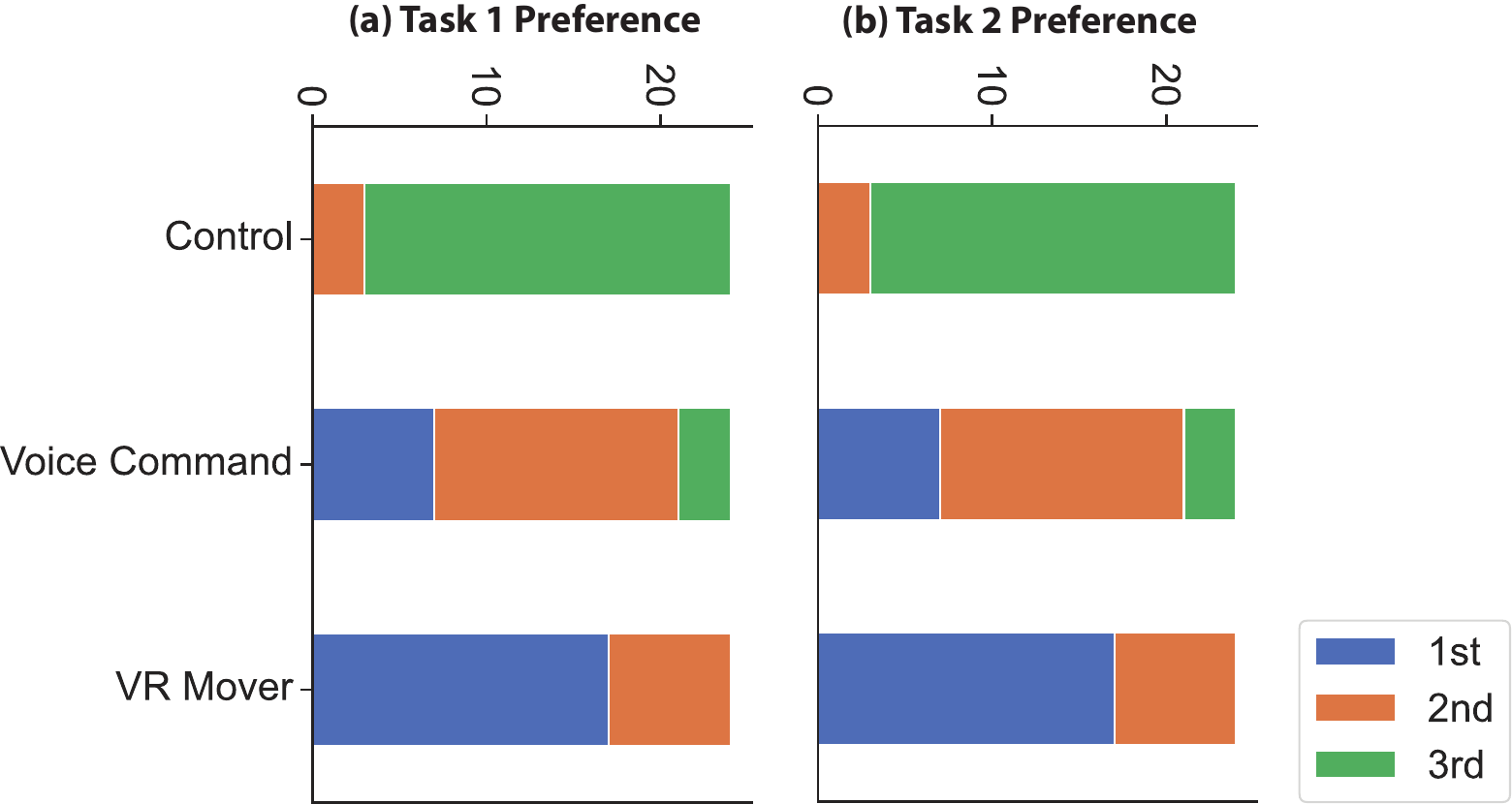}
  \caption{The preference ranking for (a) Task 1 and (b) Task 2.}
  \Description{The preference ranking for Task 1 and Task 2. The number of votes is presented in Sec 6.8.}
  \label{fig:ranking}
\end{figure}

At the end of the experiment, the users were asked to rank \textit{Control}, \textit{Voice Command} and \textit{VR Mover} for both Task 1 (\autoref{fig:ranking}\textit{a}) and Task 2 (\autoref{fig:ranking}\textit{b}). For Task 1's first choice, 2, 7, and 15 participants picked \textit{Control}, \textit{Voice Command} and VR Mover, respectively. For the second choice, the distribution is 1, 15, and 8. For Task 2, 
For Task 2, 17 participants picked \textit{VR Mover} as their first choice while the remaining 7 picked \textit{Voice Command}. For the second choice, 3, 14, and 7 participants picked Control, \textit{Voice Command} and \textit{VR Mover} respectively. Thus, generally, \textit{VR Mover} is favored by most participants, followed by \textit{Voice Command} and typical \textit{Control}. 

\subsection{Qualitative Feedback}
In addition, we have recorded the audio in the structured interview. The interview records were transcribed for coding. Later in the discussion, we will present some of the qualitative feedback of the users.

\section{Discussion}

Inspired by how humans convey spatial manipulation to each another in the real world, \textit{VR Mover} aims to mimic this behavior in VR. The \textit{VR Mover} can listen to the user's instructions and take note of their gestural cues (through pointing and lining) to determine how to assist them with coarse object placement. 
Combined with the results presented earlier, we believe that the \textit{VR Mover} is a significant contribution as a virtual object manipulation interface in terms of better performance, user experience and naturalness, while generating less arm fatigue and workload.

\subsection{Performance}
In physical reality, it is often better when there are others who can help you move objects. This is of course, particularly true when there are multiple objects to move. \textit{VR Mover} provides an efficient interface to indicate several objects (explicitly by pointing or implicitly with description) and the target(s) which may be given separately per object. As such, \textit{VR Mover} is expected to be able to complete multi-object manipulation in a short period. 
Some of the users (N=4) have commented that VR Mover is convenient, 
"I can manipulate multiple objects at a time, but others have to do it one by one. It is more convenient and quicker." (P18)

Objectively, the expectation and feedback that \textit{VR Mover} is more convenient and efficient for multi-object placement is reflected in the manipulation time result of Task 1; there, we can observe a significant difference between \textit{VR Mover} and the other two interfaces, \textit{Voice Command} and \textit{Control}. For coarse placement, \textit{VR Mover} can reduce at least $40\%$ of the time on average compared to \textit{Voice Command} and \textit{Control}. However, as indicated by the weaker effect when compared to \textit{Voice Command} and \textit{Control} for fine manipulation time, \textit{VR Mover} may not particularly assist in the fine-tuning process. This is expected as currently it is not embedded with a method for fine-tuning the placement of objects. Still, the result shows that the more efficient coarse placement can help the user to complete the task sooner. 

On the other hand, it should be noted that \textit{VR Mover} will not particularly benefit a single mid-air object scenario as a larger portion of the manipulation requires the classical gizmos $+$ virtual hand combination to rotate and vertically move in order to fit the object to the final mid-air target. However, it can be argued that mid-air placement is not necessarily common for an object manipulation scenario (e.g. the sandbox environment in Task 2). Regardless, it can be inferred that \textit{VR Mover} is best at supporting the user's performance when multiple objects are involved.

\subsection{Arm Fatigue}
When moving an object, a typical object manipulation interface will require a user to complete all the necessary actions for the entire manipulation process. On the other hand, \textit{VR Mover} can help with coarse placement, and then the user can take over to complete the fine-tuning. For the situation where there are multiple objects, the users can perform asynchronous multi-object manipulation. In addition, it should be noted that instructing \textit{VR Mover} often involves a simple pointing gesture, which should require relatively little effort from the user. As some users (N=3) has pointing out, \textit{VR Mover} is,
"more relaxing, like having a conversation."(P20)

The fact that \textit{VR Mover} induces less physical strain on the user is reflected in several places. Foremost, we can observe that \textit{VR Mover} when compared to \textit{Control} has significantly less total hand movement for Task 1B which involves multiple objects. When compared to \textit{Voice Command}, which also supports pointing, we also see this significant reduction. This can be explained by the fact that \textit{VR Mover} is better at handling multi-object manipulation.
Further, Borg C10 which measures the exertion of the users is observed to have a significant difference for \textit{VR Mover} when compared with \textit{Voice Command} and \textit{Control}. The physical subscale of NASA-TLX also shows similar significance; this indicates that besides the hand movement result, the users themselves also feel that the physical workload is less when using \textit{VR Mover}.

Similar to our discussion on performance, it should be noted that when the manipulation involves the \textit{VR Mover} less, the reduction of fatigue should similarly be reduced. This expectation can be reflected by the hand moving distance in Task 1A where \textit{VR Mover} group is virtually similar to the \textit{Control} group. However as indicated by the Borg C10 scores and the physical NASA-TLX subscale which takes into account both Task 1 and Task 2, we can see that, generally, the \textit{VR Mover} should reduce the arm fatigue of the user. 

\subsection{Workload}
Workload is an important consideration for any interface. If the workload is lower, the user will be able to use the interface for a longer period of time. We used the classical NASA-TLX to evaluate the overall workload for a technique. As reported, the mean NASA-TLX score of \textit{VR Mover} significantly lowered compared to \textit{Voice Command} and \textit{Control}. We have already mentioned the physical workload when discussing the reduced arm fatigue of \textit{VR Mover}. In the following, we will discuss other subscales in NASA-TLX. 

Foremost, we can observe that the mental demand of \textit{VR Mover} is significantly lowered compared to \textit{Voice Command} and \textit{Control}. It is believed that this lower mental demand is linked with asynchronous multi-object manipulation and coarse-to-fine design. As discussed earlier, a human's visual working memory tends to group objects together to process and manipulate visual information in a coarse-to-fine manner. By allowing users to manipulate multiple objects together and place them first on a general placement and then later finetune, \textit{VR Mover} fits the cognitive process of a human's visual working memory. Thus, the lower mental demand and the considerable effect size can be indicative of validation of this belief. 

When compared to \textit{Voice Command}, another important advantage of \textit{VR Mover} is that the user need not memorize the grammar structure and keywords. This is reflected by several users (N=6) with comments such as "(\textit{VR Mover}) is the best: no grammar requirement and quick, convenient"(P23) and that \textit{Voice Command} "must use a specific word,"(P6) %=(P30) 
and "need to remember the syntax"(P7).

For other subscales (temporal, performance, effort, and frustration), \textit{VR Mover} has significance when compared to \textit{Control} but not \textit{Voice Command}. Overall, the existing evidence might suggest that an LLM-based interface can enhance a traditional object manipulation interface by reducing the workload.

\subsection{User Experience}
User experience is an important aspect of any interface \cite{kocaballi2019understanding}; improving user experience is a useful contribution to an interface. As shown in the result section, \textit{VR Mover} achieves a significantly higher overall score in UEQ-S. In addition, \textit{VR Mover} also has a higher score for the hedonic and practical measure in UEQ-S. This may indicate that the design that echoes how we convey spatial manipulation in real life is both enjoyable and practical. Some of the users have commented that the interface is "very fun to play", "very interesting" and "amazing".
The preference ranking of \textit{VR Mover} is also encouraging for an LLM-based interface. For both Task 1 and Task 2, the majority of the users pick \textit{VR Mover} as the first choice. Together, the result shows that \textit{VR Mover} is a competitive interface for object manipulation. A few users (N=3) also commented on the freedom they feel when using the interface “Users have more freedom of expressing themselves."(P20)

\subsection{Naturalness}
A natural user interface should be intuitive to use; that is, the user is able to quickly tell how to interact via the interface \cite{mcewan2014natural}. This kind of interface is ideal because the user can dedicate less time to the "entry level" learning curve. As mentioned before, how \textit{VR Mover} aims to achieve this is to mimic a real mover that can understand natural spatial manipulation communication. Qualitatively, it seems that we have achieved our goal, some of the users (N=3) have commented on the intuitiveness of \textit{VR Mover}
“It is intuitive to instruct the system, I tell it what I want and it will automatically do it for me”(P18),
and that it feels like a conversation (N=3), “It’s like talking to a person, which makes it easy to use.”(P20)
The SUS score also seems to reflect this as \textit{VR Mover} is significantly higher compared to \textit{Control}. The two-item SUS is particularly suitable for reporting ease of use as its two questions are "I thought this software technique was easy to use" and "I found the software very cumbersome/awkward to use". Inversely, one user has commented that \textit{Control} technique for them has a "steep learning curve."(P10)%=(P34)

It was hoped that an intuitive interface would in turn improve presence. It is known that controller transparency (the user forgets about the interface) has an impact on the feeling of presence \cite{carroll2009creativity}. However, in our experiment, \textit{VR Mover} did not show significance when compared with both \textit{Voice Command} and \textit{Control}. It is believed that this could be caused by the design that the user needs to press "A" when "pointing". Thus, the user still needs to remember to press the button to signal the point or area of interest, interrupting the feeling of presence.

\subsection{Interactive Behavior}
As discussed, one of our beliefs is that users would prefer to follow a coarse-to-fine process for object manipulation. The user will instruct the \textit{VR Mover} to move the object to a general location and then finetune it via gizmos or virtual hands. During our experiments, we have observed that this is indeed quite a common behavior. Some of the users (N=4) have explicitly described this process during the interview, "When you ask someone to do the thing for you, afterward, you may still want to adjust it by yourself."(P10) %=(P34)
"(I) do the rough work in (\textit{VR Mover}) then switch to hand for precise control."(P6) %=(P30)
"(\textit{VR Mover}) is the best. Firstly use it to put things into a rough position, then use hand control for fine-tuning."(P13)%=(P37)

\section{Limitation and Future Works}
\textit{VR Mover} has several limitation. For one, it is currently not optimal for mid-air object manipulation as it is difficult to specify where to place in mid-air. Drawing a line from the ground may help decide a 3D point in the air, but it is not as straightforward as pointing directly on a surface. Second, pointing requires the user to press the A button to indicate, while in daily life, people tend to use fingers to point at interest without extra actions. A more faithful interface for pointing may further improve the user experience and effectiveness of \textit{VR Mover}.

The current scene modelling , using a bounding box to describe the objects to the LLM, is another limitation. The LLM cannot iteratively predict and perform collision checks. Moreover, the concave surface of an object cannot be handled, e.g. inserting a book into a bookshelf. Thus, improving collision checking can be a fruitful direction.
Another limitation of this work is not using visual prompting to the LLM for the sake of performance (response time). However, we believe that VR Mover can achieve better manipulation capabilities with an all-rounded visual prompt, like the ones shown in \cite{wang2024chat2layout}. Therefore, it may be worth exploring how to incorporate visual prompts while maintaining a reasonable response time. Last, the stochastic nature of the LLM \cite{Torre24} introduces varieties in the response even though given the same set of prompts. Our prompt engineering has limited the variance, but due to the nature of LLM, complete reproducibility cannot be guaranteed.

\section{Conclusion}
Inspired by how mover, in reality, can assist us in moving objects and that we have a natural capacity to convey spatial manipulation to each other, \textit{VR Mover} is an LLM-based interface that supports object manipulation. The previous interface generally requires the users to perform all the actions to reach a specific outcome; on the other hand, \textit{VR Mover} uses a user- and spatial-aware LLM to assist in the object manipulation based on the instruction of the user. The instruction need not be clear or precise, so long it is reasonably understandable.  Generally, the proposed interface follows a coarse-to-fine manipulation procedure where the user uses \textit{VR Mover} for general placement of objects and later uses a classical interface such as Gizmos or virtual hand for final adjustment. In order to realize \textit{VR Mover}, we have investigated several modules that understand the perception and interaction of the user. With them, the LLM is able to be aware of the perception of the user and the surroundings be aware such that the instruction can be well executed. 

To evaluate \textit{VR Mover}, we have conducted a user study involving performance tasks and a creative sandbox environment. Results show that compared to classical interfaces with only gizmos and virtual hand control, \textit{VR Mover} is able to be more performative for multi-object manipulation. In addition, generally, \textit{VR Mover} has better user experience, usability, and less workload and fatigue. A similar result is shown when comparing with the \textit{Voice Command} LLM-removed variant of VR Mover, which shows the importance of an LLM that can understand the natural instruction of the user. It is believed that our work provides insight for future LLM-based interfaces.

\bibliographystyle{ACM-Reference-Format}
\bibliography{main}

\newpage
\appendix

\section{Details of Task Two and User Results}
\subsection{The Menu}
\begin{figure}[htbp]
    \centering
    \includegraphics[width=0.8\textwidth]{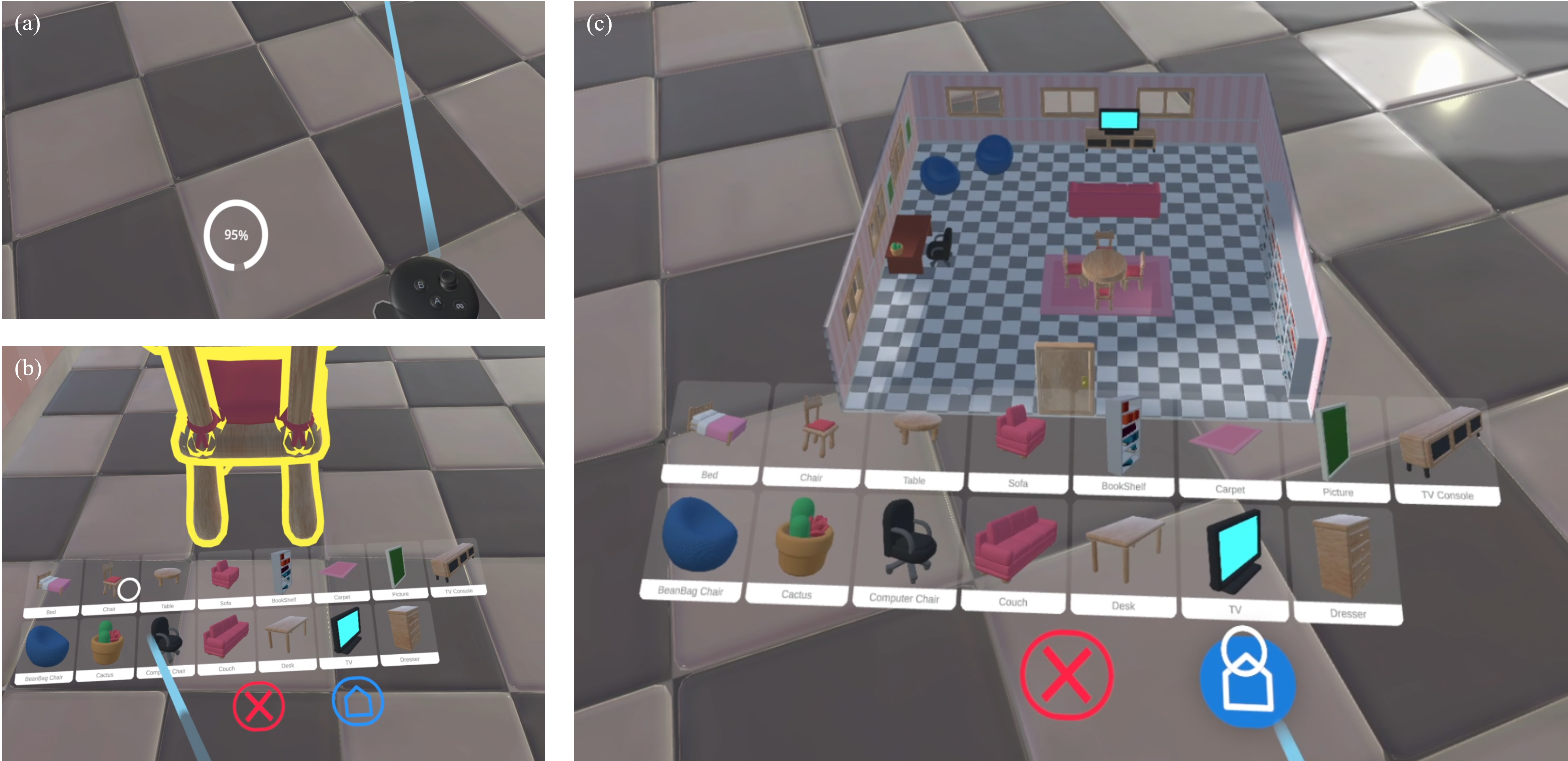}
    \caption{The menu control for task two in the user study.}
    % \Description{The menu control for the second task of the user study. (a) The progress bar shows when the user looks their head down; (b) The user can operate on the menu, a two-row collection of previews with names of the available furniture. The first row contains 8 furniture from left to right: bed, chair, table, sofa, bookshelf, carpet, picture, and TV console; while the second contains 7: beanbag chair, cactus, computer chair, couch, desk, TV, and dresser. Using the index trigger on the controller to click the thumbnail to create a new item. (c) The user hovers the hand ray on the house icon at the right bottom to show the mini-room in front of the user.}
    \label{fig:menu-panel}
\end{figure}

In the second task of the user study, the users stayed in a virtual sandbox room for 7 minutes and were asked to furnish the room with a list of furniture. The furniture prefabs are presented in the menu at \autoref{fig:menu-panel}b. To open the menu, the user must look their head down and keep 1 second. The progress bar (\autoref{fig:menu-panel}a) indicates the accumulated time of looking down and the menu then appears after the progress goes to 100\%. The users can operate the menu using the hand ray and index trigger on the control to: 1. create an item of a new furniture prefab in front of them with a single click on the thumbnail; 2. click on the close button to close the menu. In the \textit{Voice Command} and \textit{VR Mover} experiment groups, users prefer to add the furniture in the room using voice, and the menu is mainly for viewing purposes. The mini-room for task reference is accessed with the menu as well. By hovering the ray at the house icon, the mini-room pops out, as shown in \autoref{fig:menu-panel}c.

\subsection{Available Furniture Prefabs}
In total 15 types of furniture have been provided for task two, the details of them are presented in the following \autoref{tab:prefab-details}. The thumbnails and prefab IDs of the prefabs are displayed in the menu, as described above. Moreover, the prefab IDs support querying the specific one from the furniture prefabs list and API calls as parameters within the functions. Dimensions, the size of the oriented bounding box, are automatically calculated based on the 3D models used for the furniture. The dimensions are subject to change once the scale of the object instances has been altered, and are used for guiding the placement of the \textit{VR Mover}, limiting instances within the room, and identifying the in-frustum objects. The description is generated by a multi-modal LLM agent given the thumbnails, and the remarks indicate the anchor point of the prefabs, for the \textit{Voice Command} and \textit{VR Mover} to translate or rotate the object instances in the scene. In a sense, the prefab ID with the description provides the semantic information for the furniture, and the dimension describes the spatial properties.

\begin{table}[htbp]
    \centering
    \small
    \setlength{\tabcolsep}{4pt} 
    \begin{tabular}{|c|c|c|p{5.7cm}|p{2.8cm}|}
        \hline
        \textbf{Thumbnail} & \textbf{Prefab ID} & \textbf{Dimension} & \textbf{Description} & \textbf{Remarks} \\
        \hline
        \raisebox{-0.7\totalheight}{\includegraphics[height=1cm]{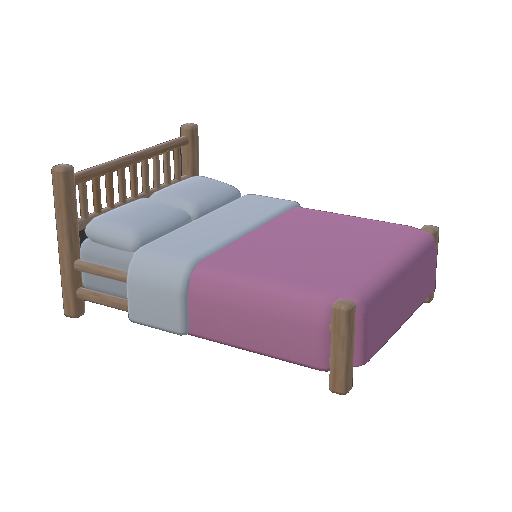}} & 
        Bed & (1.55, 0.75, 2.06) & Single bed with wooden slatted headboard and footboard. White sheets and pink blanket. Positioned with headboard facing back of scene. & Anchor: Bottom Center. \\
        \hline
        \raisebox{-0.7\totalheight}{\includegraphics[height=1cm]{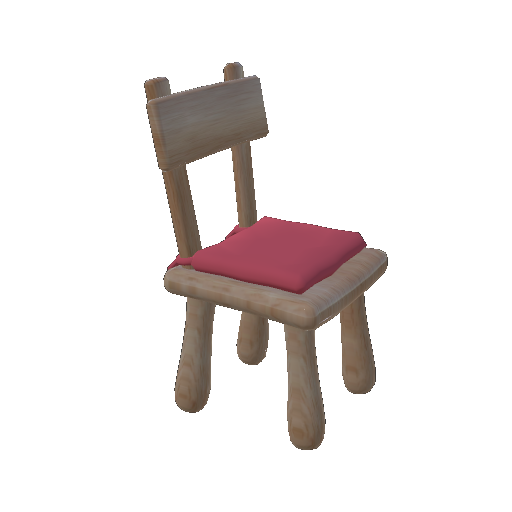}} & 
        Chair & (0.55, 1.01, 0.56) & Simple wooden chair with red seat cushion. Backrest has rounded top. Facing forward in scene. & Anchor: Bottom Center. \\
        \hline
        \raisebox{-0.7\totalheight}{\includegraphics[height=1cm]{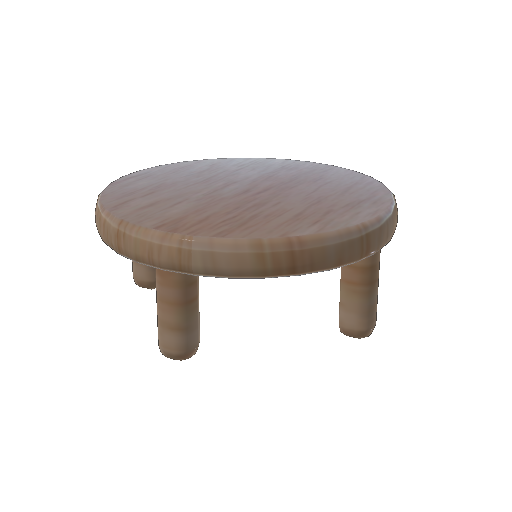}} & 
        Table & (1.49, 0.62, 0.47) & Circular coffee table with light wood frame and white top. Four short cylindrical legs. Black remote on surface. & Anchor: Bottom Center.\\
        \hline
        \raisebox{-0.7\totalheight}{\includegraphics[height=1cm]{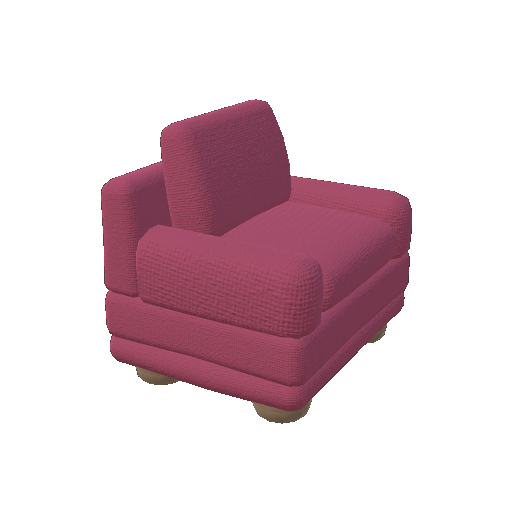}} & 
        Sofa & (1.18, 1.16, 0.96) & Pink armchair with rounded edges. Cushioned seat and backrest. Four small wheels. Facing forward in scene. & Anchor: Bottom Center. \\
        \hline
        \raisebox{-0.7\totalheight}{\includegraphics[height=1cm]{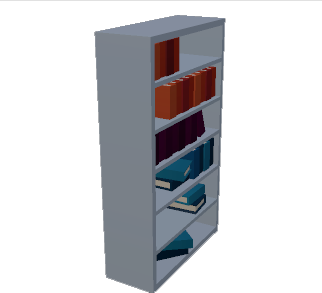}} & 
        Bookshelf & (1.17, 1.95, 0.45) & A tall white bookshelf with multiple shelves. It contains various colorful books and items arranged on different levels. The bookshelf is standing upright, facing the front of the scene. & Anchor: Bottom Center.\\
        \hline
        \raisebox{-0.7\totalheight}{\includegraphics[height=1cm]{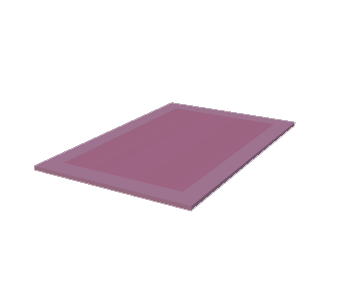}} & 
        Carpet & (1.35, 0.03, 0.90) & A simple rectangular pink rug or mat with rounded corners. It's lying flat on the floor, adding a splash of color to the room. & Anchor: Bottom Center. \\
        \hline
        \raisebox{-0.7\totalheight}{\includegraphics[height=1cm]{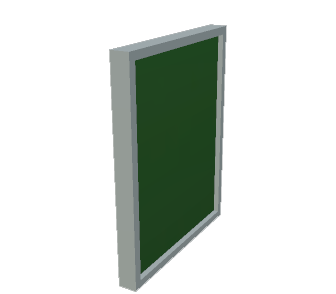}} & 
        Picture & (0.68, 0.81, 0.07) & A framed picture with a green background. The frame appears to be light-colored, possibly white or off-white. The picture is standing upright, ready to be hung on a wall or placed on a surface. & Anchor: Center of back surface.\newline Position: Place using center of picture back as reference point. \\
        \hline
        \raisebox{-0.7\totalheight}{\includegraphics[height=1cm]{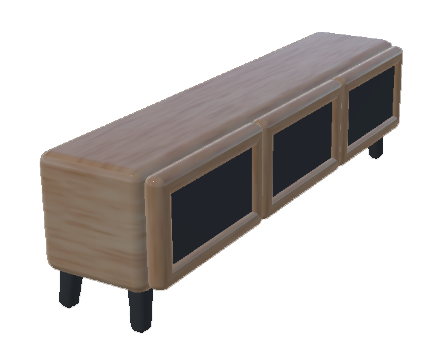}} & 
        TV Console & (2.26, 0.66, 0.53) & Long wooden sideboard with white top. Three black-paneled doors. Metal legs. Positioned lengthwise in scene. & Anchor: Bottom Center. \\
        \hline
        \raisebox{-0.7\totalheight}{\includegraphics[height=1cm]{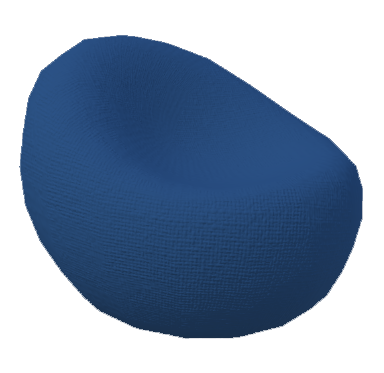}} & 
        Beanbag Chair & (0.68, 1.20, 0.63) & Curved blue egg-shaped chair with smooth contours. Single-piece design with a deep seat and rounded backrest. No visible legs or base. & Anchor: Bottom Center. \\
        \hline
        \raisebox{-0.7\totalheight}{\includegraphics[height=1cm]{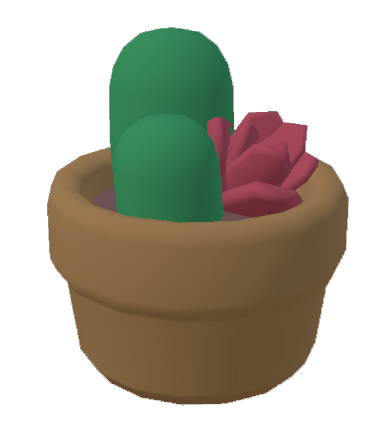}} & 
        Cactus & (0.34, 0.38, 0.34) & A pot of Cactus. & Anchor: Bottom Center. \\
        \hline
        \raisebox{-0.7\totalheight}{\includegraphics[height=1cm]{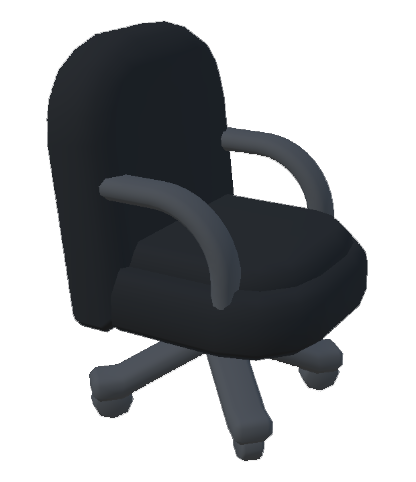}} & 
        Computer Chair & (0.69, 1.07, 0.72) & A Computer Chair with 4 wheels. & Anchor: Bottom Center.\\
        \hline
        \raisebox{-0.7\totalheight}{\includegraphics[height=1cm]{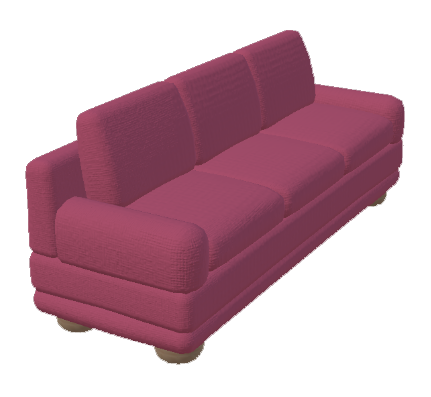}} & 
        Couch & (2.67, 1.14, 0.98) & A couch with three seats. & Anchor: Bottom Center.\\
        \hline
        \raisebox{-0.7\totalheight}{\includegraphics[height=1cm]{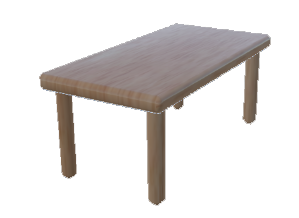}} & 
        Desk & (1.72, 0.74, 0.88) & Rectangular desk with light wood frame and white top. Four cylindrical legs. Small black accent on tabletop edge. Positioned lengthwise in scene. & Anchor: Bottom Center.\\
        \hline
        \raisebox{-0.7\totalheight}{\includegraphics[height=1cm]{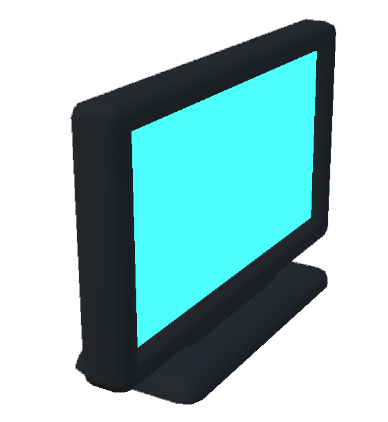}} & 
        TV & (1.28, 0.85, 0.41) & Flat-screen monitor with black frame. Cyan display. Stand at base. Facing forward in scene. & Anchor: Bottom Center.\\
        \hline
        \raisebox{-0.7\totalheight}{\includegraphics[height=1cm]{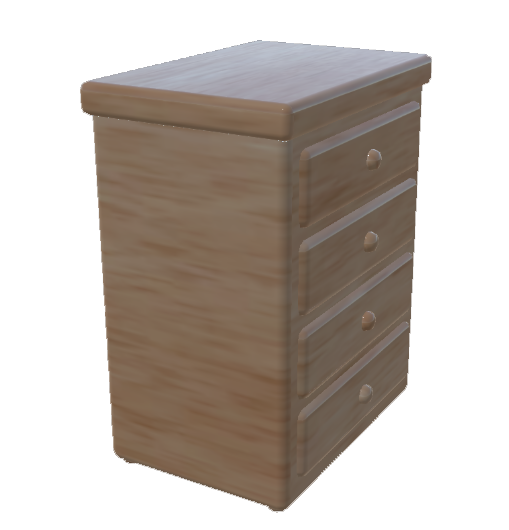}} & 
        Dresser & (0.98, 1.25, 0.72) & Wooden bedside table with four drawers. Light wood finish. Square top with slightly rounded edges. & Anchor: Bottom Center.\\
        \hline
    \end{tabular}
    \caption{Furniture prefabs}
    % \Description{TODO}
    \label{tab:prefab-details}
\end{table}

\subsection{User Built Rooms}
\begin{figure}[htbp]
    \centering
    \includegraphics[width=\textwidth]{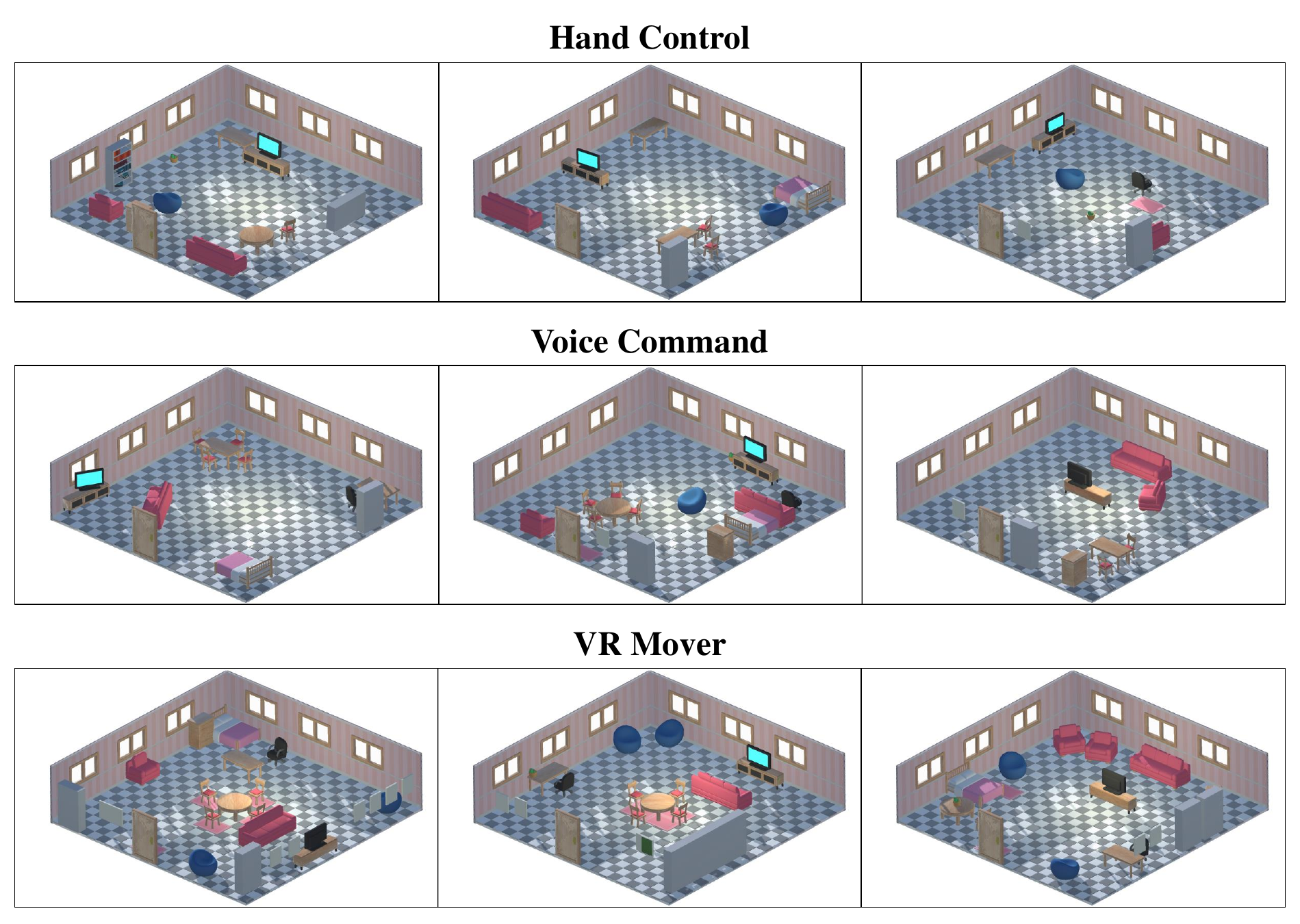}
    \caption{User built rooms via different techniques}
    % \Description{TODO}
    \label{fig:user-built-room}
\end{figure}

Here, we present the rooms built by the participants during the second task during the user study. The most rational, complete, and representative three for each technique are shown in \autoref{fig:user-built-room}. In terms of layout complexity, the number of furniture, and the alignments, it seems that the \textit{VR Mover} performs the best, the \textit{Voice Command} stays in the middle, and the \textit{Control} group performs the last. We believe that this can be reflected by how the users interact with different techniques during the study as well. During usage of the \textit{Control} and \textit{Voice Command}, participants seldom considered batch manipulations by selecting multiple objects at the same time. On the other hand, the \textit{VR Mover} internally handled the unstructured user instructions (e.g. ``Put 4 carpet touching each other under the table and chairs.'' during designing the middle room in \autoref{fig:user-built-room} \textit{VR Mover}), with the help of the user-centric data, transferred the requests into a set of structured API calls (including creating, moving, and rotating multiple objects). Each piece of furniture has its uniqueness in placement, and  \textit{Control} and \textit{Voice Command} require the user to deal with the uniqueness while the \textit{VR Mover} handles them for the users, which may explain this.

\newpage
\section{Voice Command Grammar}

The grammar structure applied to the \textit{Voice Command} in the experiments is: \texttt{\textlangle verb, subject, (direction), (unit)\textrangle }, where the ``\texttt{()}'' indicates optional input. Regular expression and mapping check each voice-transcribed-text from the user. Success command executes immediately, while the failed warns the user with reasons. The detailed explanation of the types of keywords is below:

\myListItem{\texttt{verb}} \texttt{= \{create, delete, move, rotate, scale\}} includes all the object manipulation types in the scope of this paper. For each of the \texttt{verb}, the belonging \texttt{subject}, \texttt{direction}, and \texttt{unit} could be different. The mappings are in the following.

\myListItem{\texttt{subject}} \texttt{= \{create: \textlangle prefab ID\textrangle , (delete, move, rotate, scale): \{\textlangle object name\textrangle , this\}\}} specifies the target(s) to be manipulated. The \texttt{\textlangle prefab ID\textrangle } can only follow \texttt{verb = \{create\}} for creating a new object instance in the scene. \texttt{\textlangle object name\textrangle } (for example, ``Chair (No.)2'', ``TV Console (No.)1'') is the combination of \texttt{\textlangle prefab ID\textrangle } and numbering of this type of object for clarification (system generated, displayed in the label when using the ray to hover an object), and ``\texttt{this}'' refers to all the selected objects in the current scene. Both of \texttt{\textlangle object name\textrangle } and \texttt{\textlangle prefab ID\textrangle } can follow the \texttt{verb = \{delete, move, rotate, scale\}} that manipulating the objects existing in the scene.
 
\myListItem{\texttt{direction}} \texttt{= \{move: \{left, right, up, down, forward, backward\}, rotate: \{left, right, up, down\}, scale: \{\textlangle empty\textrangle, length, width, height\}, here\}} indicates the direction for manipulating the objects for \texttt{verb = \{move, rotate, scale\}}. All the directions refer to the object(s)'s local direction and applied on one or more among the ``X'', ``Y'', or ``Z'' axis(es) with the sign: \texttt{\{move: \{left(X-), right(X+), up(Y+), down(Y-), forward(Z+), backward(Z-)\}, rotate: \{left(Y-), right(Y+), up(X+), down(X-)\}, scale: \{\textlangle empty\textrangle (X+Y+Z+), length(Z+), width(X+), height(Y+)\}\}}. ``\texttt{here}'', a special keyword, indicates a location for ``\texttt{move}'' and ``\texttt{rotate}''. It refers to the latest pointing (the same pointing system for the \textit{VR Mover}), and for ``\texttt{move}'', the system places the object(s)(specified by \texttt{subject}) to the exact position of the pointing. Similarly, ``\texttt{rotate}'' rotates the object(s), by setting the forward direction of the object(s) as the normalized vector from the object(s) to the pointing position, so that the object(s) will face to the pointing position. 

\myListItem{\texttt{unit}} \texttt{= \textlangle\textlangle number\textrangle, \textlangle units\textrangle\textrangle}, where \texttt{units = \{move: \{centimeter(s), meter(s), CM, M\}, rotate: \{degree(s)\}, scale: \textlangle empty\textrangle \}}, defines the magnitude of changes on the object. The first in the set is applied when the \texttt{\textlangle units\textrangle} is not captured in the text. The entire element can be absent, and the system uses the default value to perform corresponding actions. ``\texttt{move}'' uses 5cm as default, and ``\texttt{rotate}'' uses 30 degrees as default, while ``\texttt{scale}'' uses 1.0 as the default as it sets the absolute scaling to the object(s).

During the development and the pilot study, the voice-to-text model may insert unwanted words between the keywords, or use different methods to interpret the numerical after the word ``number'', for example, it parses ``create chair'' to ``create a chair'', ``No.5'' to ``\#5'', ``number 5'', or ``number five''. Thus, tolerances have been applied to the \textit{Voice Command}, meaningless words in between keywords are ignored, special characters are filtered out (anything out of ``[a-zA-Z0-9]'' and blank space), and the numbers in the word form are converted to digit form (only from ``one'' to ``twenty''). The participants in the user study prefer to use ``move this here'' and ``rotate this here'' with the pointing gesture in most cases for both studies. It turns out that they may have difficulties memorizing all the keywords and combinations, and ``this'' and ``here'' are more straightforward and intuitive to use compared to other rules.

%% 1. voice recognition error.. and tolerance
%% 2. actual usage of the users.

\newpage
\section{System Prompt}
The \texttt{\textlangle Available Prefabs\textrangle}, \texttt{\textlangle Room Information\textrangle}, and \texttt{\textlangle Environment Objects\textrangle} are subject to change according to the specific scene settings, and thus inserted dynamically. A real example with all the fields inserted can be found at \autoref{sec:example-IO}.

\subsection{Task One}

\begin{lstlisting}
# Summary
You are an expert virtual assistant for a unity-developed virtual object arrangement system with a strong mathematical ability. 
Your task is to:
    1. Interpret user's requirements to manipulate objects in the room
    2. Help with move/rotate/scale the object(s) to meet user's request
    3. ONLY REPLY formatted API calls for the system to parse
    
# Available APIs

1. Set the position/rotation/scale of an object:
    MOVE(string object_id, float? x = null, float? y = null, float? z = null);
    // Set the forward direction of the object in Unity style
    FORWARD(string object_id, float x = 0, float y = 0, float z = 0);
    // Set a position for the object to look at
    LOOKAT(string object_id, float? x = null, float? y = null, float? z = null);
    SCALE(string object_id, float? x = null, float? y = null, float? z = null);
2. Send a text message:
    MESSAGE(string content);
3. Send a debug text message:
    EXPLAIN(string reason);
    
# Important Factors
When placing objects, you need to consider the following factors:
1. **Physical constraints:**
    The objects have physical properties that must be considered. For example, a bookshelf will not overlap with a table or wall;
2. **User centric:**
    The objects should be placed according to the user's perspective. e.g. When the user says "a chair on the left". The chair should be placed left in his view
    *But this shouldn't break factor 1

# User Prompt Formats
The user prompt are JSONs containing following fields:
1. Player. The transform info about the player's head
2. Objects. A list of objects in the scene for you to manipulate, please note:
    - position is showing the bottom center anchor of the object, consider collision when placing objects
    - rotation of the object is represented by forward, right, and up directions
    - boundary is defined by an orientated bounding box of the object defined by center, size, and directions
3. Head Stay Frames. The frame that recording where the player's looking for a while, please note:
    - The object and environment objects are sorted, the higher weight it is, the more important it maybe
    - Infer the objects/places the player mentioned based on the frames, instead of "2. Objects"
4. Hit Points. The recorded click points by the player for indicating a place to perform tasks, please note:
    - The point is either at the environment surface or on the objects, take the normal into consideration when placing objects
    - The position IS inaccurate, do not directly use the data but consider the region. NOTE: DO NOT PUT MULTIPLE OBJECTS AT THE SAME POSITION
5. Drawing Lines. A line starting from an environment element or object indicating a direction or magnitude, please note:
    - A drawing line indicates two possible lines, one is from Start to End, another is from Start to Project. End infers to the drawing end point while Project is the project point into the environment/object of the drawing end.
    - The normal of the Start and End is the raycastHit surface of objects with name, tne End's normal is the vector from Start to End 
    - Inference if this line means a direction or magnitude from the user's request
    - The direction/magnitude might be inaccurate, try to align it with axis/environment, considering other information provided
6. User Request. Direct request of the player recognized from voice, might be inaccurate on words
7. User Request with Actions Inserted. The user request inserted with the IDs of the hit points or drawing lines, indicating what is the player talking about when doing the actions:
    - The user may say "here" / "there" / "this place" / etc. to indicate the actions.
8. Enable Actions. Telling you want API is allowed to call or not by that request 
        
# Replay Principle
## Format
Available API calls with correct syntax and no comments

## Process Steps:
1) Infer the objectiveness of the player;
2) Move / Rotate the object to the asked destination;
3) Repeat 2, 3, and 4, until request met;

**Notes:**
- Call SCALE only when explicitly required;
- Call MESSAGE only when the player asks questions, or you cannot perform the required task;
- Each line should end with ";".
- Keep two decimals;
- No math expressions in your responses.
- If the user request is so ambiguous or likely irrelevant, do nothing but reply the user using MESSAGE(string content);
- User may use hit point or/and drawing lines to indicate multiple steps to manipulate one/more object(s), 
  perform the actions step by step following the user's order, and DO CONSIDER the latest status of the object, especially your API calls just before!!

## Object Prefabs and Room Information

**Coordination System (Right-handed):**
x-axis: right;
y-axis: up;
z-axis: forward;

**Available Prefabs:**
The list of available objects for you to manipulate and alter
<prefabs_info>

**Room Information:**
<room_info>

**Environment Objects:**
The list of environment objects in the room you can see but cannot alter
<env_objects>
\end{lstlisting}

\newpage
\subsection{Task Two}
\begin{lstlisting}
# Summary
You are an expert and creative room design assistant for a unity-developed virtual room arrangement system with a strong mathematical ability. 
Your task is to:
    1. Interpret user's requirements to manipulating objects in the room
    2. Help with design of the room with your creativity and insights
    3. Fulfill the user's needs and provide suggestions (if asked)
    4. ONLY REPLY formatted API calls for the system to parse

# Available APIs

1. Create an instance based on the prefab ID:
    CREATE(string prefab_id);
2. Set the position/rotation/scale of an object:
    MOVE(string object_id, float? x = null, float? y = null, float? z = null);
    // Set the forward direction of the object in Unity style
    FORWARD(string object_id, float x = 0, float y = 0, float z = 0);
    // Set a position for the object to look at
    LOOKAT(string object_id, float? x = null, float? y = null, float? z = null);
    SCALE(string object_id, float? x = null, float? y = null, float? z = null);
3. Delete an object by its ID:
    DELETE(string object_id);
4. Send a text message:
    MESSAGE(string content);
5. Send a debug text message:
    EXPLAIN(string reason);

* object ID can be set as \"crt\" to refer to the object that called in CREATE / MOVE / ROTATE / LOOKAT / SCALE just before

# Important Factors
When placing objects, you need to consider the following factors:
1. **Physical constraints:**
    The objects have physical properties that must be considered. For example, a bookshelf will not overlap with a table or wall; Or a picture will not fly midair
2. **User centric:**
    The objects should be placed according to the user's perspective. e.g. When the user says "a chair on the left". The chair should be placed left in his view
    *But this shouldn't break factor 1
3. **Spatial relation:**
    You should consider the relations between objects. e.g. when placing chairs and a table. The chairs should be reasonably close to the table and face the table;
    *But this shouldn't break factor 2

# User Prompt Formats
The user prompt are JSONs containing following fields:
1. Player. The transform info about the player's head
2. Objects. A list of objects in the scene for you to manipulate, please note:
    - position is showing the bottom center anchor of the object except the picture, consider collision when placing objects
    - rotation of the object is represented by forward, right, and up directions
    - boundary is defined by an orientated bounding box of the object defined by center, size, and directions
3. Head Stay Frames. The frame that recording where the player's looking for a while, please note:
    - The object and environment objects are sorted, the higher weight it is, the more important it maybe
    - Infer the objects/places the player mentioned based on the frames, instead of "2. Objects"
4. Hit Points. The recorded click points by the player for indicating a place to perform tasks, please note:
    - The point is either at the environment surface or on the objects, take the normal into consideration when placing objects
    - The position IS inaccurate, do not directly use the data but consider the region. NOTE: DO NOT PUT MULTIPLE OBJECTS AT THE SAME POSITION
5. Drawing Lines. A line starting from an environment element or object indicating a direction or magnitude, please note:
    - A drawing line indicates two possible lines, one is from Start to End, another is from Start to Project. End infers to the drawing end point while Project is the project point into the environment/object of the drawing end.
    - The normal of the Start and End is the raycastHit surface of objects with name, tne End's normal is the vector from Start to End 
    - Infer if this line means a direction or magnitude from the user's request
    - The direction/magnitude might be inaccurate, try to align it with axis/environment, considering other information provided
6. User Request. Direct request of the player recognized from voice, might be inaccurate on words
7. User Request with Actions Inserted. The user request inserted with the IDs of the hit points or drawing lines, indicating what is the player talking about when doing the actions:
    - The user may say "here" / "there" / "this place" / etc. to indicate the actions.
8. Enable Actions. Telling you want API is allowed to call or not by that request 
        
# Replay Principle
## Format
Available API calls with correct syntax and no comments

## Process Steps:
1) Infer the objectiveness of the player;
2) Create a new object (if requested);
3) Move the object to the reasonable/asked destination;
4) Rotate or make the object look at in the reasonable/asked direction;
5) Repeat 2, 3, and 4, until request met;

## Principles:
1. Objects cannot in the air, placed at the same position, or overlapped with each other
    - Simply move objects to the same or close position will cause SERIOUS overlapping problem
    - Use the bounding box's orientation and dimension to check the collision is a must
2. Objects completely stay inside the room
    - There are environment objects describing the room, walls, floor, and ceiling is the boundary of the room
    - To ensure staying in the room entirely, collision check via the bounding box is a must
3. Orientation of the objects must be proper
    - Objects align with the wall has the same forward direction with the wall
    - Object face to another object when there is a relationship (e.g., table vs. chair, TV vs. couch/sofa)

**Notes:**
- Call SCALE only when explicitly required;
- Call MESSAGE only when the player asks questions, or you cannot perform the required task;
- Each line should end with ";".
- Keep two decimals;
- No math expressions in your responses.
- If the user request is so ambiguous or likely irrelevant, do nothing but reply the user using MESSAGE(string content);

## Object Prefabs and Room Information

**Coordination System (Right-handed):**
x-axis: right;
y-axis: up;
z-axis: forward;

**Available Prefabs:**
The list of available objects for you to manipulate and alter
<prefabs_info>

**Room Information:**
<room_info>

**Environment Objects:**
The list of environment objects in the room you can see but cannot alter
<env_objects>
\end{lstlisting}

\newpage
\section{Example Input and Output of the LLM}\label{sec:example-IO}
\begin{figure}[h]
    \centering
    \includegraphics[width=\textwidth]{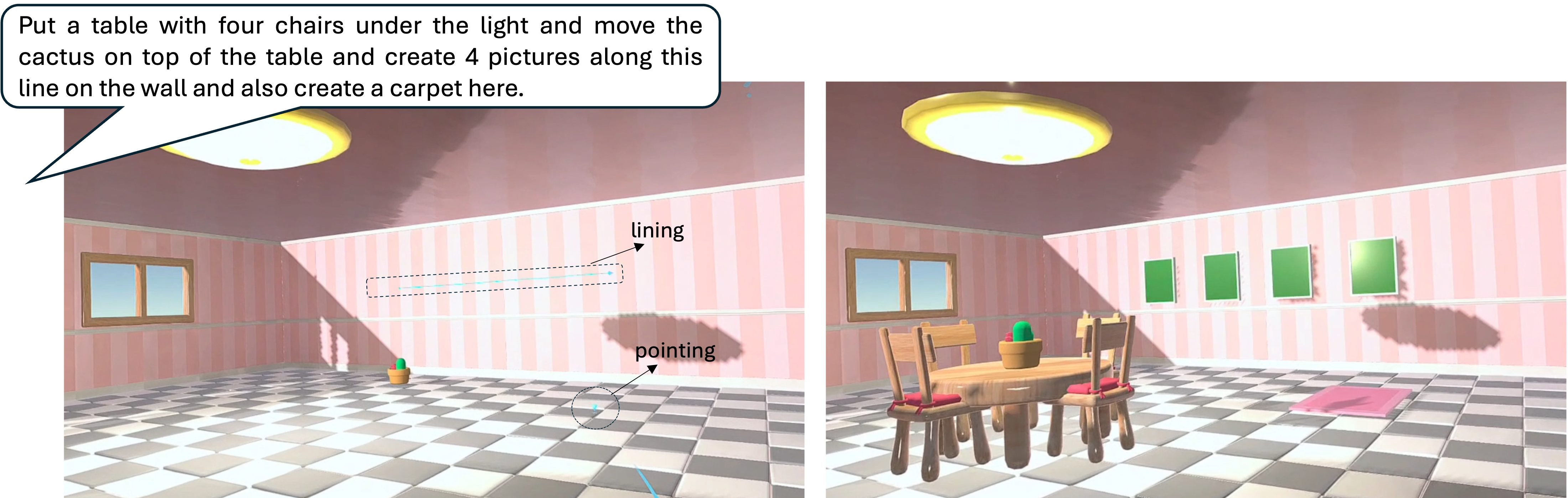}
    \caption{The task two scene before (left) and after (right) the user request has been processed by the \textit{VR Mover}.}
    % \Description{TODO}
    \label{fig:before-after-exmaple}
\end{figure}

The example presents a scenario for manipulating the object in the environment of task two during the user study. It includes manipulating the existing object in the scene, using gesture cues (pointing and lining) to indicate positions, and creating new objects and placing them. The \autoref{fig:before-after-exmaple} illustrates the scene appearance before and after the user's request has been processed by the \textit{VR Mover}, with the display of the pointing and lining. The \textbf{input} for sending to the background LLM GPT-4o is the combination of the system prompt (\ref{subsec:system-prompt}) and user prompt (\ref{subsec:user-prompt}), while the output of the LLM is the assistant prompt (\ref{subsec:assistant-prompt}).

\subsection{System Prompt}\label{subsec:system-prompt}
\begin{lstlisting}
... <Omitted fixed prompt partition, same to the System Prompt in Task Two at previous section>

**Available Prefabs:**
The list of available objects for you to manipulate and alter
{
  "prefabs": [
    {
      "prefab_id": "Bed",
      "description": "Single bed with wooden slatted headboard and footboard. White sheets and pink blanket. Positioned with headboard facing back of scene.",
      "remarks": "Anchor: Bottom Center.",
      "dimensions": {
        "x": "1.55",
        "y": "0.75",
        "z": "2.06"
      }
    },
    {
      "prefab_id": "Chair",
      "description": "Simple wooden chair with red seat cushion. Backrest has rounded top. Facing forward in scene.",
      "remarks": "Anchor: Bottom Center.",
      "dimensions": {
        "x": "0.55",
        "y": "1.01",
        "z": "0.56"
      }
    },
    {
      "prefab_id": "Table",
      "description": "Circular coffee table with light wood frame and white top. Four short cylindrical legs. Black remote on surface.",
      "remarks": "Anchor: Bottom Center.",
      "dimensions": {
        "x": "1.49",
        "y": "0.62",
        "z": "1.47"
      }
    },
    {
      "prefab_id": "Sofa",
      "description": "Pink armchair with rounded edges. Cushioned seat and backrest. Four small wheels. Facing forward in scene.",
      "remarks": "Anchor: Bottom Center",
      "dimensions": {
        "x": "1.18",
        "y": "1.16",
        "z": "0.96"
      }
    },
    {
      "prefab_id": "BookShelf",
      "description": "A tall white bookshelf with multiple shelves. It contains various colorful books and items arranged on different levels. The bookshelf is standing upright, facing the front of the scene.",
      "remarks": "Anchor: Bottom Center.",
      "dimensions": {
        "x": "1.17",
        "y": "1.95",
        "z": "0.45"
      }
    },
    {
      "prefab_id": "Carpet",
      "description": "A simple rectangular pink rug or mat with rounded corners. It's lying flat on the floor, adding a splash of color to the room.",
      "remarks": "Anchor: Bottom Center.",
      "dimensions": {
        "x": "1.35",
        "y": "0.03",
        "z": "0.90"
      }
    },
    {
      "prefab_id": "Picture",
      "description": "A framed picture with a green background. The frame appears to be light-colored, possibly white or off-white. The picture is standing upright, ready to be hung on a wall or placed on a surface.",
      "remarks": "Anchor: Center of back surface.\nPosition: Place using center of picture back as reference point.",
      "dimensions": {
        "x": "0.68",
        "y": "0.81",
        "z": "0.07"
      }
    },
    {
      "prefab_id": "TV Console",
      "description": "Long wooden sideboard with white top. Three black-paneled doors. Metal legs. Positioned lengthwise in scene.",
      "remarks": "Anchor: Bottom Center.",
      "dimensions": {
        "x": "2.26",
        "y": "0.66",
        "z": "0.53"
      }
    },
    {
      "prefab_id": "BeanBag Chair",
      "description": "Curved blue egg-shaped chair with smooth contours. Single-piece design with a deep seat and rounded backrest. No visible legs or base.",
      "remarks": "Anchor: Bottom Center.",
      "dimensions": {
        "x": "0.68",
        "y": "1.20",
        "z": "0.63"
      }
    },
    {
      "prefab_id": "Cactus",
      "description": "A pot of Cactus",
      "remarks": "Anchor: Bottom Center.",
      "dimensions": {
        "x": "0.34",
        "y": "0.38",
        "z": "0.34"
      }
    },
    {
      "prefab_id": "Computer Chair",
      "description": "A Computer Chair with 4 wheels",
      "remarks": "Anchor: Bottom Center.",
      "dimensions": {
        "x": "0.69",
        "y": "1.07",
        "z": "0.72"
      }
    },
    {
      "prefab_id": "Couch",
      "description": "A couch with three seats",
      "remarks": "Anchor: Bottom Center.",
      "dimensions": {
        "x": "2.67",
        "y": "1.14",
        "z": "0.98"
      }
    },
    {
      "prefab_id": "Desk",
      "description": " Rectangular desk with light wood frame and white top. Four cylindrical legs. Small black accent on tabletop edge. Positioned lengthwise in scene.",
      "remarks": "Anchor: Bottom Center.",
      "dimensions": {
        "x": "1.72",
        "y": "0.74",
        "z": "0.88"
      }
    },
    {
      "prefab_id": "TV",
      "description": "Flat-screen monitor with black frame. Cyan display. Stand at base. Facing forward in scene.",
      "remarks": "Anchor: Bottom Center.",
      "dimensions": {
        "x": "1.28",
        "y": "0.85",
        "z": "0.41"
      }
    },
    {
      "prefab_id": "Dresser",
      "description": "Wooden bedside table with four drawers. Light wood finish. Square top with slightly rounded edges.",
      "remarks": "",
      "dimensions": {
        "x": "0.98",
        "y": "1.25",
        "z": "0.72"
      }
    }
  ]
}

**Room Information:**
Room Center: (4.49, 0.05, 4.39)
Room Dimensions: (11.00, 3.00, 11.00)


**Environment Objects:**
The list of environment objects in the room you can see but cannot alter
[
    {
        "name": "Window4",
        "boundary": {
            "Central": {
                "x": "1.00",
                "y": "1.59",
                "z": "10.00"
            },
            "Size": {
                "x": "1.81",
                "y": "1.10",
                "z": "0.18"
            },
            "Forward": {
                "x": "0.00",
                "y": "0.00",
                "z": "1.00"
            },
            "Up": {
                "x": "0.00",
                "y": "1.00",
                "z": "0.00"
            },
            "Right": {
                "x": "1.00",
                "y": "0.00",
                "z": "0.00"
            }
        }
    },
    {
        "name": "Window2",
        "boundary": {
            "Central": {
                "x": "-1.00",
                "y": "1.59",
                "z": "4.00"
            },
            "Size": {
                "x": "0.18",
                "y": "1.10",
                "z": "1.81"
            },
            "Forward": {
                "x": "0.00",
                "y": "0.00",
                "z": "1.00"
            },
            "Up": {
                "x": "0.00",
                "y": "1.00",
                "z": "0.00"
            },
            "Right": {
                "x": "1.00",
                "y": "0.00",
                "z": "0.00"
            }
        }
    },
    {
        "name": "Window6",
        "boundary": {
            "Central": {
                "x": "7.00",
                "y": "1.59",
                "z": "10.00"
            },
            "Size": {
                "x": "1.81",
                "y": "1.10",
                "z": "0.18"
            },
            "Forward": {
                "x": "0.00",
                "y": "0.00",
                "z": "1.00"
            },
            "Up": {
                "x": "0.00",
                "y": "1.00",
                "z": "0.00"
            },
            "Right": {
                "x": "1.00",
                "y": "0.00",
                "z": "0.00"
            }
        }
    },
    {
        "name": "Window3",
        "boundary": {
            "Central": {
                "x": "-1.00",
                "y": "1.59",
                "z": "7.00"
            },
            "Size": {
                "x": "0.18",
                "y": "1.10",
                "z": "1.81"
            },
            "Forward": {
                "x": "0.00",
                "y": "0.00",
                "z": "1.00"
            },
            "Up": {
                "x": "0.00",
                "y": "1.00",
                "z": "0.00"
            },
            "Right": {
                "x": "1.00",
                "y": "0.00",
                "z": "0.00"
            }
        }
    },
    {
        "name": "Corner4",
        "boundary": {
            "Central": {
                "x": "9.85",
                "y": "0.09",
                "z": "-0.88"
            },
            "Size": {
                "x": "0.10",
                "y": "0.10",
                "z": "0.10"
            },
            "Forward": {
                "x": "0.00",
                "y": "0.00",
                "z": "1.00"
            },
            "Up": {
                "x": "0.00",
                "y": "1.00",
                "z": "0.00"
            },
            "Right": {
                "x": "1.00",
                "y": "0.00",
                "z": "0.00"
            }
        }
    },
    {
        "name": "Corner2",
        "boundary": {
            "Central": {
                "x": "-0.88",
                "y": "0.09",
                "z": "9.86"
            },
            "Size": {
                "x": "0.10",
                "y": "0.10",
                "z": "0.10"
            },
            "Forward": {
                "x": "0.00",
                "y": "0.00",
                "z": "1.00"
            },
            "Up": {
                "x": "0.00",
                "y": "1.00",
                "z": "0.00"
            },
            "Right": {
                "x": "1.00",
                "y": "0.00",
                "z": "0.00"
            }
        }
    },
    {
        "name": "Corner3",
        "boundary": {
            "Central": {
                "x": "9.85",
                "y": "0.09",
                "z": "9.86"
            },
            "Size": {
                "x": "0.10",
                "y": "0.10",
                "z": "0.10"
            },
            "Forward": {
                "x": "0.00",
                "y": "0.00",
                "z": "1.00"
            },
            "Up": {
                "x": "0.00",
                "y": "1.00",
                "z": "0.00"
            },
            "Right": {
                "x": "1.00",
                "y": "0.00",
                "z": "0.00"
            }
        }
    },
    {
        "name": "Window5",
        "boundary": {
            "Central": {
                "x": "4.00",
                "y": "1.59",
                "z": "10.00"
            },
            "Size": {
                "x": "1.81",
                "y": "1.10",
                "z": "0.18"
            },
            "Forward": {
                "x": "0.00",
                "y": "0.00",
                "z": "1.00"
            },
            "Up": {
                "x": "0.00",
                "y": "1.00",
                "z": "0.00"
            },
            "Right": {
                "x": "1.00",
                "y": "0.00",
                "z": "0.00"
            }
        }
    },
    {
        "name": "Wall_Z_Positive",
        "boundary": {
            "Central": {
                "x": "4.49",
                "y": "1.29",
                "z": "-0.91"
            },
            "Size": {
                "x": "10.75",
                "y": "2.60",
                "z": "0.00"
            },
            "Forward": {
                "x": "0.00",
                "y": "0.00",
                "z": "1.00"
            },
            "Up": {
                "x": "0.00",
                "y": "1.00",
                "z": "0.00"
            },
            "Right": {
                "x": "1.00",
                "y": "0.00",
                "z": "0.00"
            }
        }
    },
    {
        "name": "LightCeiling",
        "boundary": {
            "Central": {
                "x": "5.00",
                "y": "2.52",
                "z": "5.00"
            },
            "Size": {
                "x": "1.63",
                "y": "0.27",
                "z": "1.63"
            },
            "Forward": {
                "x": "0.00",
                "y": "0.00",
                "z": "1.00"
            },
            "Up": {
                "x": "0.00",
                "y": "1.00",
                "z": "0.00"
            },
            "Right": {
                "x": "1.00",
                "y": "0.00",
                "z": "0.00"
            }
        }
    },
    {
        "name": "Window1",
        "boundary": {
            "Central": {
                "x": "-1.00",
                "y": "1.59",
                "z": "1.00"
            },
            "Size": {
                "x": "0.18",
                "y": "1.10",
                "z": "1.81"
            },
            "Forward": {
                "x": "0.00",
                "y": "0.00",
                "z": "1.00"
            },
            "Up": {
                "x": "0.00",
                "y": "1.00",
                "z": "0.00"
            },
            "Right": {
                "x": "1.00",
                "y": "0.00",
                "z": "0.00"
            }
        }
    },
    {
        "name": "Door",
        "boundary": {
            "Central": {
                "x": "4.30",
                "y": "1.14",
                "z": "-0.99"
            },
            "Size": {
                "x": "1.32",
                "y": "2.20",
                "z": "0.29"
            },
            "Forward": {
                "x": "0.00",
                "y": "0.00",
                "z": "1.00"
            },
            "Up": {
                "x": "0.00",
                "y": "1.00",
                "z": "0.00"
            },
            "Right": {
                "x": "1.00",
                "y": "0.00",
                "z": "0.00"
            }
        }
    },
    {
        "name": "Wall_X_Negative",
        "boundary": {
            "Central": {
                "x": "9.94",
                "y": "1.28",
                "z": "4.53"
            },
            "Size": {
                "x": "0.00",
                "y": "2.60",
                "z": "10.75"
            },
            "Forward": {
                "x": "0.00",
                "y": "0.00",
                "z": "1.00"
            },
            "Up": {
                "x": "0.00",
                "y": "1.00",
                "z": "0.00"
            },
            "Right": {
                "x": "1.00",
                "y": "0.00",
                "z": "0.00"
            }
        }
    },
    {
        "name": "Wall_Z_Negative",
        "boundary": {
            "Central": {
                "x": "4.49",
                "y": "1.28",
                "z": "9.92"
            },
            "Size": {
                "x": "10.75",
                "y": "2.60",
                "z": "0.00"
            },
            "Forward": {
                "x": "0.00",
                "y": "0.00",
                "z": "1.00"
            },
            "Up": {
                "x": "0.00",
                "y": "1.00",
                "z": "0.00"
            },
            "Right": {
                "x": "1.00",
                "y": "0.00",
                "z": "0.00"
            }
        }
    },
    {
        "name": "Floor",
        "boundary": {
            "Central": {
                "x": "4.49",
                "y": "0.05",
                "z": "4.39"
            },
            "Size": {
                "x": "10.74",
                "y": "0.00",
                "z": "10.74"
            },
            "Forward": {
                "x": "0.00",
                "y": "0.00",
                "z": "1.00"
            },
            "Up": {
                "x": "0.00",
                "y": "1.00",
                "z": "0.00"
            },
            "Right": {
                "x": "1.00",
                "y": "0.00",
                "z": "0.00"
            }
        }
    },
    {
        "name": "Corner1",
        "boundary": {
            "Central": {
                "x": "-0.88",
                "y": "0.09",
                "z": "-0.89"
            },
            "Size": {
                "x": "0.10",
                "y": "0.10",
                "z": "0.10"
            },
            "Forward": {
                "x": "0.00",
                "y": "0.00",
                "z": "1.00"
            },
            "Up": {
                "x": "0.00",
                "y": "1.00",
                "z": "0.00"
            },
            "Right": {
                "x": "1.00",
                "y": "0.00",
                "z": "0.00"
            }
        }
    },
    {
        "name": "Ceiling",
        "boundary": {
            "Central": {
                "x": "4.49",
                "y": "2.59",
                "z": "4.39"
            },
            "Size": {
                "x": "10.74",
                "y": "0.00",
                "z": "10.74"
            },
            "Forward": {
                "x": "0.00",
                "y": "0.00",
                "z": "1.00"
            },
            "Up": {
                "x": "0.00",
                "y": "1.00",
                "z": "0.00"
            },
            "Right": {
                "x": "1.00",
                "y": "0.00",
                "z": "0.00"
            }
        }
    },
    {
        "name": "Wall_X_Positive",
        "boundary": {
            "Central": {
                "x": "-0.91",
                "y": "1.28",
                "z": "4.53"
            },
            "Size": {
                "x": "0.00",
                "y": "2.60",
                "z": "10.75"
            },
            "Forward": {
                "x": "0.00",
                "y": "0.00",
                "z": "1.00"
            },
            "Up": {
                "x": "0.00",
                "y": "1.00",
                "z": "0.00"
            },
            "Right": {
                "x": "1.00",
                "y": "0.00",
                "z": "0.00"
            }
        }
    }
]
\end{lstlisting}

\newpage
\subsection{User Prompt}\label{subsec:user-prompt}
\begin{lstlisting}
{
  "player": {
    "position": {
      "x": "2.03",
      "y": "1.18",
      "z": "1.44"
    },
    "forward": {
      "x": "0.93",
      "y": "0.06",
      "z": "0.36"
    },
    "right": {
      "x": "0.37",
      "y": "-0.07",
      "z": "-0.93"
    }
  },
  "objects": [
    {
      "object_id": "-23780",
      "object_name": "Cactus",
      "position": {
        "x": "8.71",
        "y": "0.05",
        "z": "6.20"
      },
      "scale": {
        "x": "1.00",
        "y": "1.00",
        "z": "1.00"
      },
      "boundary": {
        "Central": {
          "x": "8.71",
          "y": "0.24",
          "z": "6.20"
        },
        "Size": {
          "x": "0.34",
          "y": "0.38",
          "z": "0.34"
        },
        "Forward": {
          "x": "-0.07",
          "y": "0.00",
          "z": "1.00"
        },
        "Up": {
          "x": "0.00",
          "y": "1.00",
          "z": "0.00"
        },
        "Right": {
          "x": "1.00",
          "y": "0.00",
          "z": "0.07"
        }
      }
    }
  ],
  "head_stay_frames": [
    {
      "Stay Duration": 11.1600657,
      "Speak words": "",
      "In Frustum Objects ID": [
        {
          "Object": "-23780",
          "Weight": 163
        }
      ],
      "In Frustum Environment Objects ID": [
        {
          "Object": "LightCeiling",
          "Weight": 159
        },
        {
          "Object": "Corner3",
          "Weight": 156
        },
        {
          "Object": "Window6",
          "Weight": 155
        },
        {
          "Object": "Wall_X_Negative",
          "Weight": 151
        },
        {
          "Object": "Corner4",
          "Weight": 136
        },
        {
          "Object": "Window5",
          "Weight": 103
        },
        {
          "Object": "Door",
          "Weight": 95
        },
        {
          "Object": "Wall_Z_Negative",
          "Weight": 78
        },
        {
          "Object": "Ceiling",
          "Weight": 75
        },
        {
          "Object": "Floor",
          "Weight": 74
        },
        {
          "Object": "Wall_Z_Positive",
          "Weight": 69
        }
      ]
    },
    {
      "Stay Duration": 22.1203156,
      "Speak words": "put a table with four chairs under the light and move the cactus on top of the table and create four pictures along this line on the wall and also create a carpet here ",
      "In Frustum Objects ID": [
        {
          "Object": "-23780",
          "Weight": 333
        }
      ],
      "In Frustum Environment Objects ID": [
        {
          "Object": "Corner3",
          "Weight": 328
        },
        {
          "Object": "Window6",
          "Weight": 307
        },
        {
          "Object": "LightCeiling",
          "Weight": 290
        },
        {
          "Object": "Wall_X_Negative",
          "Weight": 172
        },
        {
          "Object": "Wall_Z_Negative",
          "Weight": 163
        },
        {
          "Object": "Window5",
          "Weight": 149
        },
        {
          "Object": "Ceiling",
          "Weight": 86
        },
        {
          "Object": "Floor",
          "Weight": 85
        }
      ]
    }
  ],
  "hit_points": [
    {
      "hit_id": "h0",
      "object": "Floor",
      "position": {
        "x": "7.54",
        "y": "0.05",
        "z": "2.99"
      },
      "normal": {
        "x": "0.00",
        "y": "1.00",
        "z": "0.00"
      }
    }
  ],
  "drawing_lines": [
    {
      "Id": "d0",
      "Start": {
        "object": "Wall_X_Negative",
        "position": {
          "x": "9.94",
          "y": "1.52",
          "z": "7.01"
        },
        "normal": {
          "x": "-1.00",
          "y": "0.00",
          "z": "0.00"
        }
      },
      "End": {
        "object": "End point",
        "position": {
          "x": "11.45",
          "y": "1.65",
          "z": "3.71"
        },
        "normal": {
          "x": "1.51",
          "y": "0.13",
          "z": "-3.29"
        }
      },
      "Project": {
        "object": "Wall_X_Negative",
        "position": {
          "x": "9.94",
          "y": "1.52",
          "z": "7.01"
        },
        "normal": {
          "x": "-1.00",
          "y": "0.00",
          "z": "0.00"
        }
      }
    }
  ],
  "user_request": "Put a table with four chairs under the light and move the cactus on top of the table and create 4 pictures along this line on the wall and also create a carpet here. ",
  "user_request_with_actions_inserted": "put a table with four chairs under the light and move the cactus on top of the table [<d0>start] and create four pictures along [<d0>end] this line on the wall and also create a [<h0>] carpet here ",
  "enabled_actions": "All the actions are available",
  "step_explain": "Debugging disabled, do not call EXPLAIN(string message); !"
}
\end{lstlisting}

\newpage
\subsection{Assistant Prompt}\label{subsec:assistant-prompt}
\begin{lstlisting}
CREATE("Table");
MOVE("crt", 5.00, 0.05, 5.00);
CREATE("Chair");
MOVE("crt", 5.00, 0.05, 5.75);
LOOKAT("crt", x=5.00, z=5.00);
CREATE("Chair");
MOVE("crt", 5.75, 0.05, 5.00);
LOOKAT("crt", x=5.00, z=5.00);
CREATE("Chair");
MOVE("crt", 5.00, 0.05, 4.25);
LOOKAT("crt", x=5.00, z=5.00);
CREATE("Chair");
MOVE("crt", 4.25, 0.05, 5.00);
LOOKAT("crt", x=5.00, z=5.00);
MOVE("-23780", 5.00, 0.67, 5.00);
CREATE("Picture");
MOVE("crt", 9.94, 1.52, 7.01);
FORWARD("crt", x=-1);
CREATE("Picture");
MOVE("crt", 10.44, 1.52, 5.76);
FORWARD("crt", x=-1);
CREATE("Picture");
MOVE("crt", 10.94, 1.52, 4.51);
FORWARD("crt", x=-1);
CREATE("Picture");
MOVE("crt", 11.44, 1.52, 3.26);
FORWARD("crt", x=-1);
CREATE("Carpet");
MOVE("crt", 7.54, 0.05, 2.99);
\end{lstlisting}

\end{document}